\documentclass[sigconf, nonacm]{acmart}
\usepackage{graphicx, amsfonts}
\usepackage{color, multirow, array, footnote, subfigure, bigints}
\usepackage{paralist} 
\usepackage{enumitem}
\usepackage[symbol]{footmisc}
\usepackage{filecontents}
\usepackage{amsfonts} 
\DeclareMathSymbol{\shortminus}{\mathbin}{AMSa}{"39}

\newcommand\vldbdoi{XX.XX/XXX.XX}
\newcommand\vldbpages{XXX-XXX}
\newcommand\vldbvolume{14}
\newcommand\vldbissue{1}
\newcommand\vldbyear{2020}
\newcommand\vldbauthors{\authors}
\newcommand\vldbtitle{\shorttitle} 
\newcommand\vldbavailabilityurl{URL_TO_YOUR_ARTIFACTS}
\newcommand\vldbpagestyle{plain}

\begin{document}

\title{Closing the B-tree vs.~LSM-tree Write Amplification Gap on Modern Storage Hardware with Built-in Transparent Compression}

\author{Yifan Qiao}
\affiliation{%
  \institution{Rensselaer Polytechnic Institute}
  \streetaddress{}
  \city{Troy}
  \state{USA}
  \postcode{}
}
\email{qiaoy2@rpi.edu}

\author{Xubin Chen}
\affiliation{%
  \institution{Google Inc.}
  \streetaddress{}
  \city{Seattle}
  \country{USA}
}
\email{xubin.chen@hotmail.com}

\author{Ning Zheng}
\affiliation{%
  \institution{ScaleFlux Inc.}
  \city{San Jose}
  \country{USA}
}
\email{ning.zheng@scaleflux.com}

\author{Jiangpeng Li}
\affiliation{%
  \institution{ScaleFlux Inc.}
  \city{San Jose}
  \country{USA}
}
\email{jiangpeng.li@scaleflux.com}

\author{Yang Liu}
\affiliation{%
  \institution{ScaleFlux Inc.}
  \city{San Jose}
  \country{USA}
}
\email{yang.liu@scaleflux.com}

\author{Tong Zhang}
\affiliation{%
  \institution{Rensselaer Polytechnic Institute}
  \city{Troy}
  \country{USA}
}

\email{zhangt4@rpi.edu}


\begin{abstract}
This paper studies the design of B-tree that can take full advantage of modern storage hardware with built-in transparent compression.  
Recent years have witnessed significant interest in applying log-structured merge tree~(LSM-tree) as an alternative to B-tree. The current consensus is that, compared with B-tree, LSM-tree has distinct advantages in terms of storage space efficiency and write amplification. This paper argues that one should revisit this belief upon the arrival of storage hardware with built-in transparent compression. Advanced storage appliances~(e.g., all-flash array) and emerging computational storage drives perform hardware-based lossless data compression, transparent to OS and user applications. Beyond straightforwardly reducing the physical storage cost difference between B-tree and LSM-tree, such modern storage hardware brings new opportunities to innovate B-tree implementation in order to largely reduce its write amplification. As the first step to explore the potential, this paper presents three simple design techniques~(i.e., deterministic page shadowing, localized page modification logging, and sparse redo logging)  that can leverage such modern storage hardware to significantly reduce the B-tree write amplification. We implemented these design techniques and carried out experiments on a commercial storage drive with built-in transparent compression. The results show that the proposed design techniques can reduce the B-tree write amplification by over 10$\times$. Compared with RocksDB (a popular key-value store built upon LSM-tree), the implemented B-tree can achieve similar or even smaller write amplification and physical storage space usage. 
\end{abstract}
\maketitle

\renewcommand{\thefootnote}{\fnsymbol{footnote}}
\pagestyle{\vldbpagestyle}
\begingroup\small\noindent\raggedright\textbf{PVLDB Reference Format:}\\
\vldbauthors. \vldbtitle. PVLDB, \vldbvolume(\vldbissue): \vldbpages, \vldbyear.\\
\href{https://doi.org/\vldbdoi}{doi:\vldbdoi}
\endgroup
\begingroup
\renewcommand\thefootnote{}\footnote{\noindent
This work is licensed under the Creative Commons BY-NC-ND 4.0 International License. Visit \url{https://creativecommons.org/licenses/by-nc-nd/4.0/} to view a copy of this license. For any use beyond those covered by this license, obtain permission by emailing \href{mailto:info@vldb.org}{info@vldb.org}. Copyright is held by the owner/author(s). Publication rights licensed to the VLDB Endowment. \\
\raggedright Proceedings of the VLDB Endowment, Vol. \vldbvolume, No. \vldbissue\ %
ISSN 2150-8097. \\
\href{https://doi.org/\vldbdoi}{doi:\vldbdoi} \\
}\addtocounter{footnote}{-1}\endgroup

\ifdefempty{\vldbavailabilityurl}{}{
\vspace{.3cm}
\begingroup\small\noindent\raggedright\textbf{PVLDB Artifact Availability:}\\
The source code, data, and/or other artifacts have been made available at \url{\vldbavailabilityurl}.
\endgroup
}

\section{Introduction}
\label{sec:introduction}
This paper presents a B-tree design solution optimized for a growing family of commercial data storage hardware that internally carry out high-speed hardware-based lossless data compression, transparent to the host OS and user applications. 
Modern all-flash array products~(e.g., Dell EMC PowerMAX~\cite{DellECM-PowerMax-link}, HPE Nimble Storage~\cite{HPE-Nimble-link}, and Pure Storage FlashBlade~\cite{Pure-Storage-link}) almost always come with the built-in hardware-based transparent compression capability. Commercial solid-state storage drives with built-in transparent compression are emerging~(e.g., computational storage drive from ScaleFlux~\cite{ScaleFlux-link} and Nytro SSD from Seagate~\cite{Seagate-FMS-19}). Moreover, Cloud vendors have started to integrate hardware-based compression capability into their storage infrastructure~(e.g., Microsoft Corsia~\cite{Microsoft-acceleration-19} and AWS Graviton2~\cite{AWS-graviton-link}), leading to imminent arrival of cloud-based storage hardware with built-in transparent compression. By using dedicated hardware compression engines, such storage hardware support high-throughput data compression and decompression at very low latency and zero host CPU overhead. 

As the most widely used indexing data structure, B-tree~\cite{graefe2011modern} powers almost all the relational database management systems~(RDBMs) and hence plays a crucial role in determining the performance and efficiency of modern information technology infrastructure. Recently, log-structured merge tree~(LSM-tree)~\cite{o1996log} has attracted significant interest as a contender to B-tree, mainly because its data structure could enable higher storage space usage efficiency and lower write amplification than B-tree. The arrival of storage hardware with built-in transparent compression could readily reduce or even eliminate the gap between B-tree and LSM-tree in terms of storage space usage efficiency. This paper shows that such storage hardware can also be leveraged to significantly reduce B-tree write amplification, which closes the write amplification gap with LSM-tree as well. The key is to exploit the fact that such storage hardware allows data management software employ {\it sparse data structure} without sacrificing the true physical storage cost. In particular, when running on such storage hardware, data management software could leave 4KB LBA~(logical block address) blocks partially filled with real data or even completely empty, without wasting the physical storage space usage. Intuitively, the feasibility of employing sparse data structure creates a new spectrum of design space for innovating data management systems~\cite{Zheng-Hotstorage-20}.


This paper shows that B-tree could leverage sparse data structure enabled by such storage hardware to largely reduce its write amplification. We note that the write amplification is measured based on the amount of data being written to the physical storage media (i.e., after in-storage compression), other than the amount of data being written to the logical storage space (i.e., before in-storage compression). 
In particular, this paper presents three simple yet effective design techniques: (1) {\it deterministic page shadowing} that can ensure B-tree page update atomicity without incurring extra write overhead, (2) {\it localized page modification logging} that can reduce the write amplification caused by the mismatch between the B-tree page size and the size of modified data within each page, and (3) {\it sparse redo logging} that can reduce the write amplification caused by B-tree redo logging~(or write-ahead logging). All these three techniques share the theme of appropriately increasing the storage data structure sparsity to reduce the physical write amplification, without sacrificing the physical storage space consumption. With significantly reduced write amplification, B-tree can support much higher insert/update throughput, and more readily accommodate low-cost, low-endurance NAND flash memory~(e.g., 4bits/cell QLC and even 5bits/cell PLC NAND flash memory).

Accordingly, we implemented a B-tree (called B$^\shortminus$-tree) that incorporates these three simple design techniques. We further compared it with LSM-tree and normal B-tree by using RocksDB~\cite{RocksDB-link} and WiredTiger~\cite{Wiredtiger-link} (the default storage engine of MongoDB) as representatives, respectively.
We carried out experiments on a commercial computational storage drive with built-in transparent compression~\cite{ScaleFlux-link}. The results well demonstrate the effectiveness of the proposed design techniques on reducing the B-tree write amplification. For example, under random write workloads with 128B per record, RocksDB and WiredTiger (with page size of 8KB) have write amplification of 14 and 64, respectively, while our B$^\shortminus$-tree~(with 8KB page size) has a write amplification of only 8, representing 43\% and 88\% reduction compared with RocksDB and WiredTiger, respectively. The smaller write amplification can directly translate into a higher write throughput. For example, our results show that, under random write workloads, B$^\shortminus$-tree can achieve about 85K TPS (transactions per second), whilte the TPS of RocksDB and WiredTiger is 71K and 28K, respectively. Moreover, we note that the proposed design techniques mainly confine within the I/O module of B-tree and are largely orthogonal to the core B-tree in-memory architecture and operations. As a result, it is relatively easy to incorporate the proposed design techniques into existing B-tree implementations. For example, upon a baseline B-tree implementation, we only modified/added about 1,200 LoC to incorporate the proposed three design techniques.


\section{Background}
\label{sec:background}

\subsection{B-tree Data Compression}
\label{sec:B-treecompression}
\begin{figure}[hbtp]
  \centering
  \includegraphics[width = 0.9\linewidth]{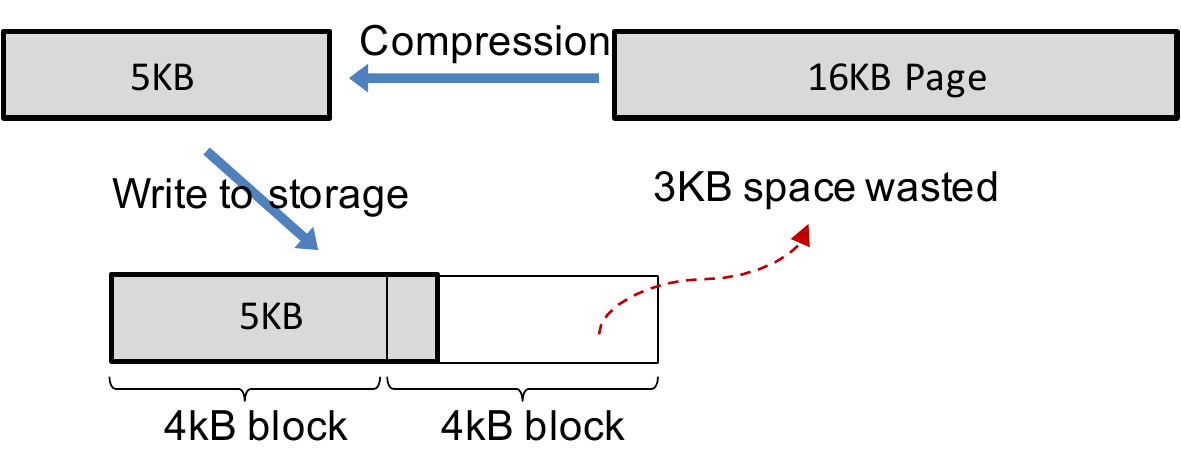}
  \caption{An example to show the storage space waste caused by 4KB-alignment constraint for B-tree page compression.}
  \label{fig:Btreewaste}
\end{figure}

B-tree manages its data storage in the unit of page (or B-tree node). To reduce data storage cost, B-tree could apply block compression algorithms (e.g., lz4~\cite{lz4-link}, zlib~\cite{zlib-link}, and ZSTD~\cite{zstd-link}) to compress each on-storage page~(e.g., the page compression feature in MySQL and MongoDB/WiredTiger). In addition to the obvious CPU overhead, B-tree page compression suffers from compression ratio loss due to the 4KB-alignment constraint, which can be explained as follows: Modern storage devices serve IO requests in the unit of 4KB LBA blocks. As a result, each B-tree page~(regardless of compressed or uncompressed) must entirely occupy one or multiple 4kB LBA blocks on the storage device (i.e., no two pages could share one LBA block on storage devices). When B-tree applies page compression, the 4KB-alignment constraint could noticeable or even significant storage space waste. This can be illustrated in Fig.~\ref{fig:Btreewaste}: Assume one 16KB B-tree page is compressed to 5KB, the compressed page must occupy two LBA blocks~(i.e., 8KB) on the storage device, wasting 3KB storage space. Therefore, due to the CPU overhead and storage space waste caused by the 4KB-alignment constraint, B-tree page compression is not widely used in production environment. Moreover, it is well-known that, under workloads with random writes, B-tree pages tend to be only 50\%$\sim$80\% full~\cite{graefe2011modern}. Hence, B-tree typically has a low storage space usage efficiency. In contrast, LSM-tree has a much more compact data structure and is free from the 4KB-alignment constraint in case of compression, which leads to a higher storage space usage efficiency than B-tree.

\subsection{In-Storage Transparent Compression}
\label{sec:CSD}
\begin{figure}[hbtp]
  \centering
  \includegraphics[width = \linewidth]{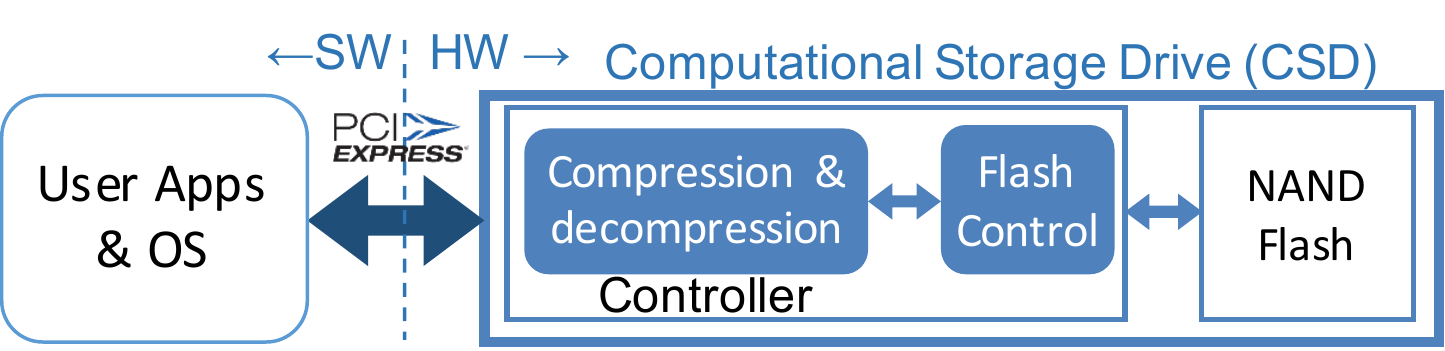}
  \caption{Illustration of a CSD with built-in transparent compression.}
  \label{fig:SSD}
\end{figure}

Fig.~\ref{fig:SSD} illustrates a computational storage drive with built-in transparent compression: Inside the computational storage drive controller chip, compression and decompression are carried out directly on the I/O path by the hardware engine, and the FTL~(flash translation layer) manages the mapping of all the variable-length compressed data blocks. Since the compression is carried out inside the storage drive, it is not subject to 4KB-alignment constraint~(i.e., all the compressed blocks are packed tightly in flash memory without any waste). 

\begin{figure}[hbtp]
  \centering
  \includegraphics[width=0.95\linewidth]{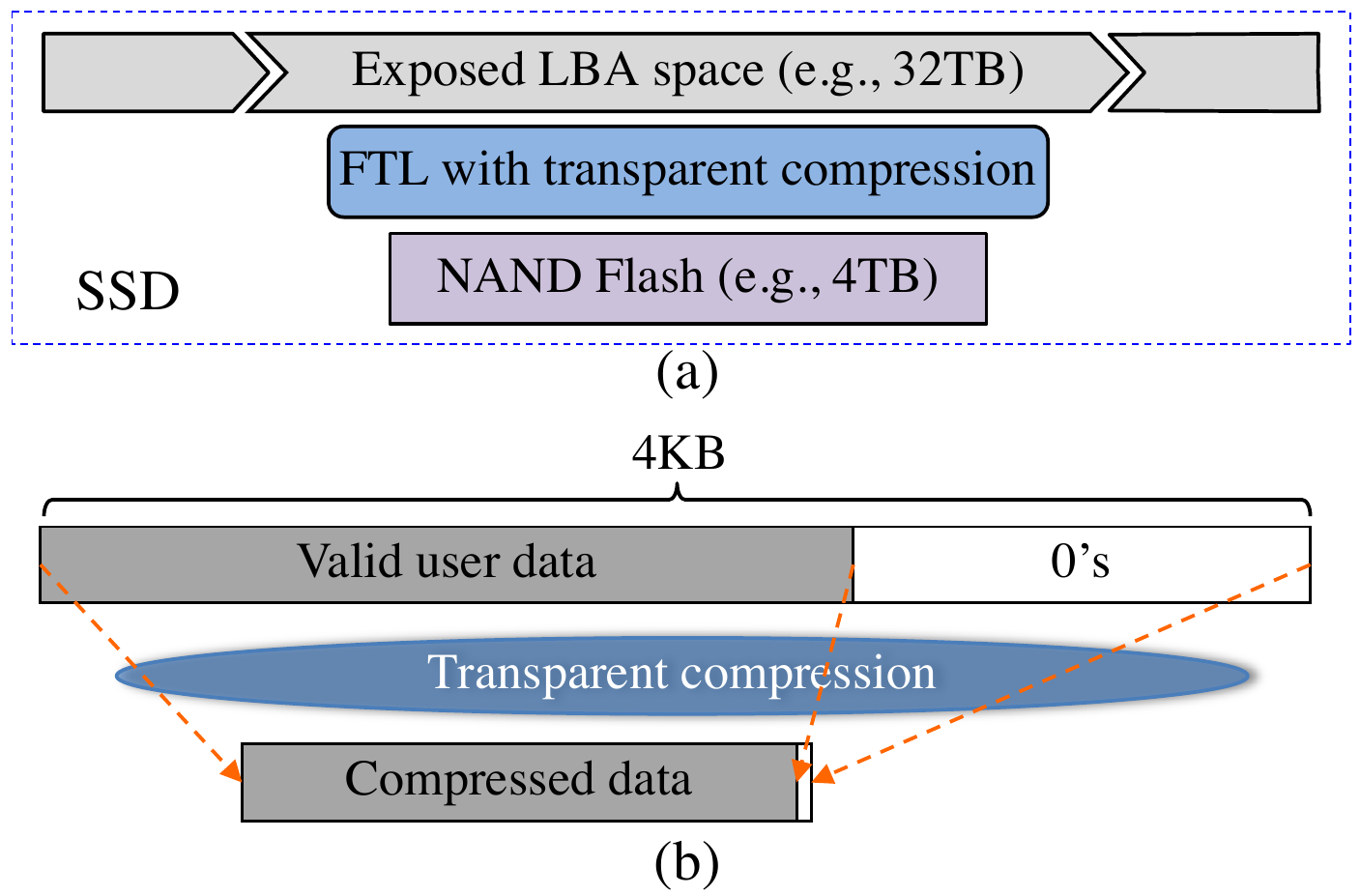}
  \caption{Illustration of the decoupled logical and physical storage space utilization efficiency enabled by storage hardware with built-in transparent compression.} \label{fig:decoupled}
\end{figure}

As illustrated in Fig.~\ref{fig:decoupled}, storage hardware with built-in transparent compression has the following two properties: (a) The storage hardware can expose an LBA space that is larger or even much larger than its internal physical storage capacity. This is conceptually similar to the thin provisioning. (b) Since certain special data patterns (e.g., all-zero or all-one) can be highly compressed, we can leave one 4KB LBA partially filled with valid data without wasting the physical storage space. These two properties essentially decouple the logical storage space utilization efficiency from the physical storage space utilization efficiency. This allows data management software systems to employ {\it sparse data structure} in the logical storage space without sacrificing the true physical storage cost, which creates a new spectrum of design space for data management systems~\cite{Zheng-Hotstorage-20}.

\subsection{B-tree vs.~LSM-tree}
\label{sec:B-tree-LSM-tree}
\begin{figure}[hbtp]
  \centering
  \includegraphics[width=0.9\linewidth]{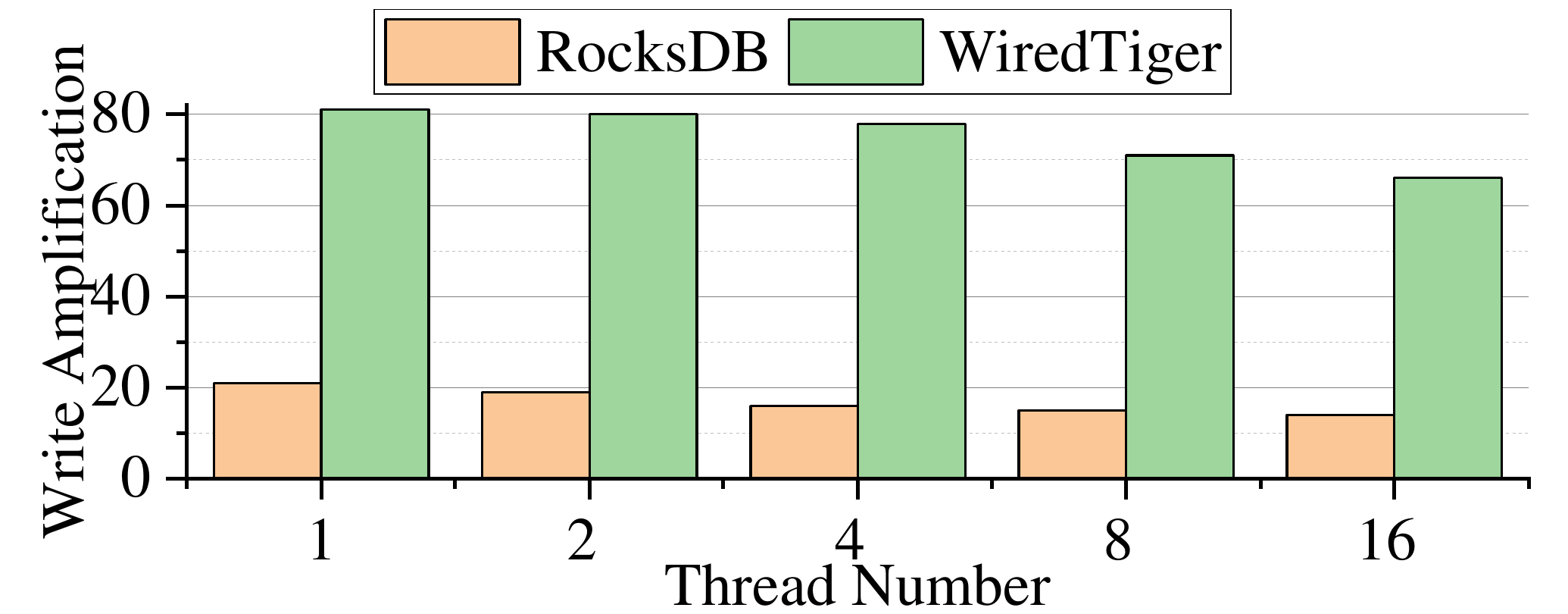}
  \caption{Measured write amplification.} \label{fig:De}
\end{figure}

As an alternative to B-tree, LSM-tree has recently received significant interest (e.g., see~\cite{yue2016building, PebblesDB-17, balmau2017triad, lu2017wisckey, ren2017slimdb, dayan2018dostoevsky, huang2019x, Luo-LSM-survey-20}) because of its advantages in terms of storage space usage and write amplification. When B-tree and LSM-tree operate on storage hardware with built-in transparent compression, their storage space usage difference may largely reduce, while LSM-tree still maintains its distinct advantage on write amplification. For the purpose of demonstration, we use RocksDB and WiredTiger (the default storage engine of MongoDB) as representatives of LSM-tree and B-tree, and carried out experiments on a 3.2TB storage drive with built-in transparent compression that was recently launched by ScaleFlux~\cite{ScaleFlux-link}. We run random write-only workloads with 128-byte record size over 150GB dataset size. For both RocksDB and WiredTiger, we disabled the application-level compression and write-ahead log (WAL), and kept all the other settings as their default value. For WiredTiger, we set its B-tree leaf page size as 8KB. Table~\ref{tab:storage} lists both the logical storage usage on the LBA space (i.e., before in-storage compression) and physical storage usage of flash memory (i.e., after in-storage compression). Since LSM-tree has a more compact data structure than B-tree, RocksDB has a smaller logical storage space usage than WiredTiger (i.e., 218GB vs.~280GB). Nevertheless, after in-storage transparent compression, WiredTiger consumes even less physical storage space than RocksDB, most likely due to the space amplification of LSM-tree. Fig.~\ref{fig:De} shows the measured write amplification under different number of client threads. We note that we measure the write amplification based on the volume of post-compression data being physically written to NAND flash memory inside the storage drive. The results show that RocksDB consistently has about 4$\times$ less write amplification than WiredTiger.
\begin{table}[htbp]
\centering
\caption{Storage space usage comparison.}
\label{tab:storage} 
\begin{tabular}{|c|c|c|}
\hline
& \multicolumn{2}{c|}{Storage space usage}\\
\cline{2-3}
&\hspace*{12pt} Logical \hspace*{12pt}& \hspace*{12pt} Physical \hspace*{12pt} \\ 
\hline
\hspace*{12pt}RocksDB\hspace*{12pt} &218GB & 129GB\\
\hline
WiredTiger &280GB & 104GB\\
\hline
\end{tabular}
\end{table}

The above results suggest that, by simply replacing normal SSDs with such modern storage hardware, we could close the physical storage cost gap between B-tree and LSM-tree, while LSM-tree still maintains its significant advantage in terms of write amplification. The goal of this work is to investigate whether we could further close the write amplification gap by appropriately modifying the B-tree implementation.

\subsection{B-tree Write Amplification}
\label{sec:B-treeWA}
We define B-tree write amplification as the ratio between the total amount of data written to the physical storage media by B-tree and the total amount of user data written into B-tree. Under current I/O interface protocols, storage devices only guarantee write atomicity over each 4KB LBA block. As a result, when the page size is larger than 4KB, B-tree must on its own ensure page write atomicity, which can be realized by using two different strategies: (i) In-place page update: Although the convenient in-place update strategy simplifies the page storage management, B-tree must accordingly use page journaling~(e.g., double-write buffer in MySQL and WAL with full-page-write in PostgreSQL) to survive partial page write failures, leading to about 2$\times$ higher write volume. (ii) Copy-on-write~(or shadowing) page update: Although copy-on-write obviates the use of page journaling and readily supports snapshot, it complicates the page storage management. Meanwhile B-tree must employ certain mechanisms~(e.g., page mapping table, or page update propagation) to keep track of the page location on the logical storage LBA space, which still incurs extra write overhead. 

Accordingly, we could classify B-tree storage write traffic into three categories: (1) logging writes~(e.g., redo/undo log) that ensure transaction atomicity and isolation, (2) page writes that persist in-memory dirty B-tree pages to storage devices, and (3) extra writes that are induced by ensuring page write atomicity~(e.g., page journaling in the case of in-place updates, or page mapping table persist in the case of page shadowing). Let $W_{log}$, $W_{pg}$, and $W_{e}$ denote the total data write amount of these three categories, and $W_{usr}$ denote the total amount of user data written into the B-tree. We can express the B-tree write amplification as
\begin{eqnarray}
    WA&=&\frac{W_{log}}{W_{usr}} + \frac{W_{pg}}{W_{usr}} + \frac{W_{e}}{W_{usr}} \nonumber \\
    &=& WA_{log}+WA_{pg}+WA_{e}.
\end{eqnarray}
When B-tree runs on storage hardware with built-in transparent compression, let $\alpha_{log}$, $\alpha_{pg}$, and $\alpha_{e}$ denote the average compression ratio of the three categories of data writes. Here we calculate the compression ratio by dividing the after-compression data volume with the before-compression data volume. Hence the compression ratio always falls into $(0,1]$, and a higher data compressibility leads to a smaller compression ratio. Therefore, the overall B-tree write amplification becomes
\begin{equation}
\label{eq:WA}
    WA = \alpha_{log}\cdot WA_{log} + 
    \alpha_{pg}\cdot WA_{pg}+\alpha_{e}\cdot WA_{e}.
\end{equation}

\section{Proposed Design Techniques}
\label{sec:proposed}
According to Eq.~(\ref{eq:WA}) above, we can reduce the B-tree write amplification by either reducing $WA_{log}$, $WA_{pg}$, and/or $WA_{e}$~(i.e., reducing the B-tree write data volumes), or reducing $\alpha_{log}$, $\alpha_{pg}$, and/or $\alpha_{e}$~(i.e., improving the B-tree write data compressibility). By applying sparse data structure enabled by storage hardware with built-in transparent compression, this section presents three design techniques to reduce the B-tree write amplification: (1) deterministic page shadowing that eliminates $WA_{e}$, (2) localized page modification logging that reduces both $WA_{pg}$ and $\alpha_{pg}$, and (3) sparse redo logging that reduces $\alpha_{log}$.

\subsection{Deterministic Page Shadowing}
\label{sec:datashadow}
In order to eliminate $WA_{e}$, B-tree should employ the principle of page shadowing instead of in-place page update. Nevertheless, in conventional implementation of page shadowing, the new on-storage location of each updated B-tree page is dynamically determined during the runtime and must be recorded/persisted by B-tree, leading to extra write overhead and management complexity. To eliminate the extra write overhead and meanwhile simplify the B-tree page storage management, we propose a technique called {\it deterministic page shadowing} as illustrated in Fig.~\ref{fig:shadowing}: 
\begin{figure}[hptb]
  \centering
  \includegraphics[width=0.85\linewidth]{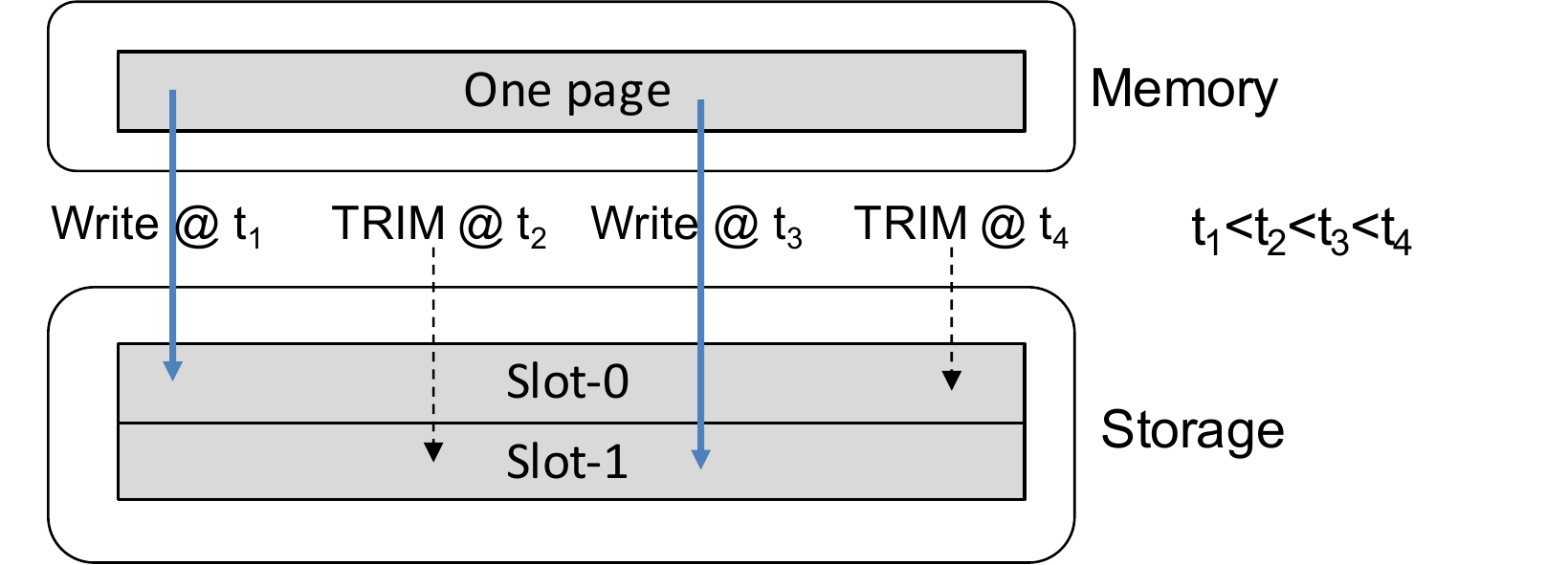}
  \caption{Illustration of deterministic page shadowing: two slots at the fixed location on the logical storage LBA space alternatively serve the memory-to-storage flush of one page.} \label{fig:shadowing}
\end{figure}
Let $l_{pg}$ denote the B-tree page size~(e.g., 8KB or 16KB). For each page, B-tree allocates $2l_{pg}$ amount of logical storage area on the LBA space and partitions it into two size-$l_{pg}$ slots (slot-0 and slot-1). For each B-tree page, the two slots at the fixed location on the logical storage LBA space serve memory-to-storage page flush alternatively in the ping-pong manner. Once a page has been successfully flushed from memory into one slot, B-tree will issue a {\it TRIM} command over the other slot. This is conceptually the same as the conventional page shadowing with the difference that the location of the shadow page is now fixed. Although B-tree occupies 2$\times$ larger logical storage area on the LBA space, only half of the storage space store valid data and the other half are trimmed (hence do not consume physical flash memory storage space). As pointed out above in Section~\ref{sec:CSD}, storage hardware with built-in transparent compression could expose a logical LBA storage space that is (much) larger than its internal physical storage capacity. Hence, such storage hardware can readily support the deterministic page shadowing. We note that deterministic page shadowing solely aims at ensuring page write atomicity without extra write overhead. To support multi-version concurrency control (MVCC), B-tree could use conventional methods such as undo logging. 

With the proposed deterministic page shadowing, B-tree uses an in-memory bitmap to keep track of the valid slot for each page. Compared with page table being used in conventional page shadowing, bitmap is much smaller and hence significantly reduces the memory usage. Moreover, B-tree does not need to persist the bitmap. In case of system re-start, B-tree can gradually rebuild the in-memory bitmap: When B-tree loads one page from storage to memory for the first time, it reads both slots from the storage device. For the trimmed slot, storage device simply returns an all-zero block, based on which B-tree can easily identify the valid slot. When B-tree reads both slots of a page, the storage device internally only fetches the valid (i.e., untrimmed) slot from the physical storage media. Hence, compared with reading one slot, reading both slots will only incur more data transfer through the PCIe interface, without any extra read latency inside the storage device. This should not be an issue as the upcoming PCIe Gen5 will support 16GB/s$\sim$32GB/s, which is significantly larger than the back-end flash memory access bandwidth inside storage devices and hence can readily accommodate the extra data transfer. In case of system crash, B-tree needs to handle the following two possible scenarios: (i) A slot is partially written before the system crash: B-tree can easily identify the partially written slot by verifying the page checksum. (ii) A slot has been successfully written but the other slot has not been trimmed before the system crash: B-tree can identify the valid slot by comparing the page LSN~(logical sequence number) of the pages on both slots. Since it is not necessary to persist the in-memory bitmap, deterministic page shadowing can completely eliminate the $\alpha_e\cdot WA_e$ component from the total B-tree write amplification.

\subsection{Localized Page Modification Logging}
\label{sec:pagelogging}
The second technique aims at reducing both $\alpha_{pg}$ and $WA_{pg}$ components in Eq.~(\ref{eq:WA}). It is motivated by a simple observation: For a B-tree page, let $\Delta$ denote the difference between its in-memory image and on-storage image. If the difference is significantly smaller than the page size (i.e., $|\Delta|<<l_{pg}$), we can largely reduce the write amplification by logging the page modification $\Delta$, instead of writing the entire in-memory page image, to the underlying storage device. This is conceptually the same as the well-known similarity-based data deduplication~\cite{aronovich2009design} and delta encoding~\cite{mogul1997potential}. Unfortunately, when B-tree runs on normal storage devices~(i.e., without built-in transparent compression), this approach is subject to significant operational overhead and hence is not practically viable: Given the 4KB block IO interface, we must coalesce multiple $\Delta$'s from different pages into one 4KB LBA block in order to materialize the write amplification reduction. To enhance the gain, we should apply the page modification logging multiple times for each page, before resetting this process to construct the up-to-date on-storage page image. Accordingly, multiple $\Delta$'s associated with the same page will spread over multiple 4KB LBA blocks on the storage device, which however will cause two problems: (1) For each page, B-tree must keep track of all its associated
$\Delta$'s and also periodically carry out background garbage collection, leading to a much higher storage management complexity. (2) To load a page from storage, B-tree has to read the existing on-storage page image and multiple $\Delta$'s from multiple non-contiguous 4KB LBA blocks. This obviously results in significant read amplification, leading to a (much) longer page load latency. Therefore, to our best knowledge, this simple design concept has not been used by real-world B-tree implementations ever reported in the open literature.

Storage hardware with built-in transparent compression for the first time makes it practically viable to implement the simple idea of page modification logging. By applying sparse data structure enabled by such storage hardware, we no longer have to coalesce multiple $\Delta$'s from different pages into the same 4KB LBA block. Leveraging the abundant logical storage LBA space, for each B-tree page, we can simply dedicate one 4KB LBA block as its modification logging space to store the $\Delta$, which is referred to as localized page modification logging. Under the 4KB IO interface, to realize the proposed page modification logging for each page, B-tree writes $D=[\Delta, {\bf O}]$~(where ${\bf O}$ represents an all-zero vector, and $|D|$ is 4KB) to the 4KB block associated with the page. Inside the storage device, all the zeros in $D$ will be compressed away and only the compressed version of $\Delta$ will be physically stored. Therefore, when serving each memory-to-storage page flush with page modification logging, we reduce $WA_{pg}$ by writing 4KB instead of $l_{pg}$ amount of data to the logical storage LBA space, and reduce the compression ratio $\alpha_{pg}$ since the written data $[\Delta, {\bf O}]$ can be highly compressed by the storage device. By dedicating one 4KB modification logging space for each B-tree page, we do not incur extra B-tree storage management complexity. The read amplification is small for two main reasons: (1) B-tree always reads only one additional 4KB LBA block. Moreover, each page and its associated 4KB logging block contiguously reside on the LBA space. Hence, in order to read both the page and its associated 4KB logging block, B-tree only issues a single read request to the storage device. (2) The storage device internally fetches very small amount of data from flash memory in order to reconstruct the 4KB LBA block $[\Delta, {\bf O}]$.  

To practically implement this simple idea, B-tree must carry out two extra operations: (1) To load a page from storage into memory, B-tree must construct the up-to-date page image based on the on-storage page image and $\Delta$. (2) To flush a page from memory to storage, B-tree must obtain $\Delta$ and accordingly decide whether it could invoke the page modification logging. To minimize the operational overhead, B-tree could apply the following strategy: Let $P_m$ and $P_s$ denote the in-memory and on-storage images of one B-tree page. We logically partition $P_m$ and $P_s$ into $k$ segments, i.e., $P_m=[P_{m,1},\cdots,P_{m,k}]$ and $P_s=[P_{s,1},\cdots,P_{s,k}]$, and $|P_{m,i}|=|P_{s,i}|$ $\forall i$ (i.e., the two segments $P_{m,i}$ and $P_{s,i}$ at the same position have the same size). For each page, B-tree keeps a $k$-bit vector $f=[f_1, \cdots, f_k]$, where $f_i$ is set to 1 if $P_{m,i}\neq P_{s,i}$. Accordingly, we construct $\Delta$ by concatenating all the in-memory segments $P_{m,i}$ with $f_i=1$. During the runtime, whenever the $i$-th segment in one in-memory page is modified, B-tree will set its corresponding $f_i$ as 1. When B-tree flushes a page from memory to storage, it first calculates the size of $\Delta$ as
\begin{equation}
    |\Delta|=\sum_{\forall i, f_i=1}|P_{m,i}|.
    \label{eq:3}
\end{equation}

We define a fixed threshold $T$ that is not larger than 4KB. If $|\Delta|\le T$, then B-tree will invoke the page modification logging, where $\Delta$ can be obtained by simple memory-copy operations. We note that the $k$-bit vector $f$ should be written together with $\Delta$ into the dedicated 4KB page modification logging block. When B-tree loads a page from storage into memory, it fetches $l_{pg}+4KB$ amount of data from the storage device, where the size-$l_{pg}$ space contains the current on-storage page image $P_s$ and the additional 4KB block contains the associated $f$ and $\Delta$. Accordingly, we could easily construct the up-to-date page image through simple memory-copy operations. For each B-tree page, the size of its $\Delta$ will monotonically increase as B-tree undergoes more write operations. Once $|\Delta|$ becomes larger than the threshold $T$, we will reset the process by flushing the entire up-to-date page to storage with $\Delta=\emptyset$ and $f$ being an all-zero vector. We note that the threshold $T$ configures the trade-off between write amplification reduction and storage space amplification: As we increase the value of $T$, we can less frequently reset the page modification logging process, leading to a smaller write amplification. Meanwhile, under a larger value of $T$, more page modifications will accumulate in the logging space and cause a larger storage cost overhead. 

\begin{figure}[b]
  \centering
  \includegraphics[width=\linewidth]{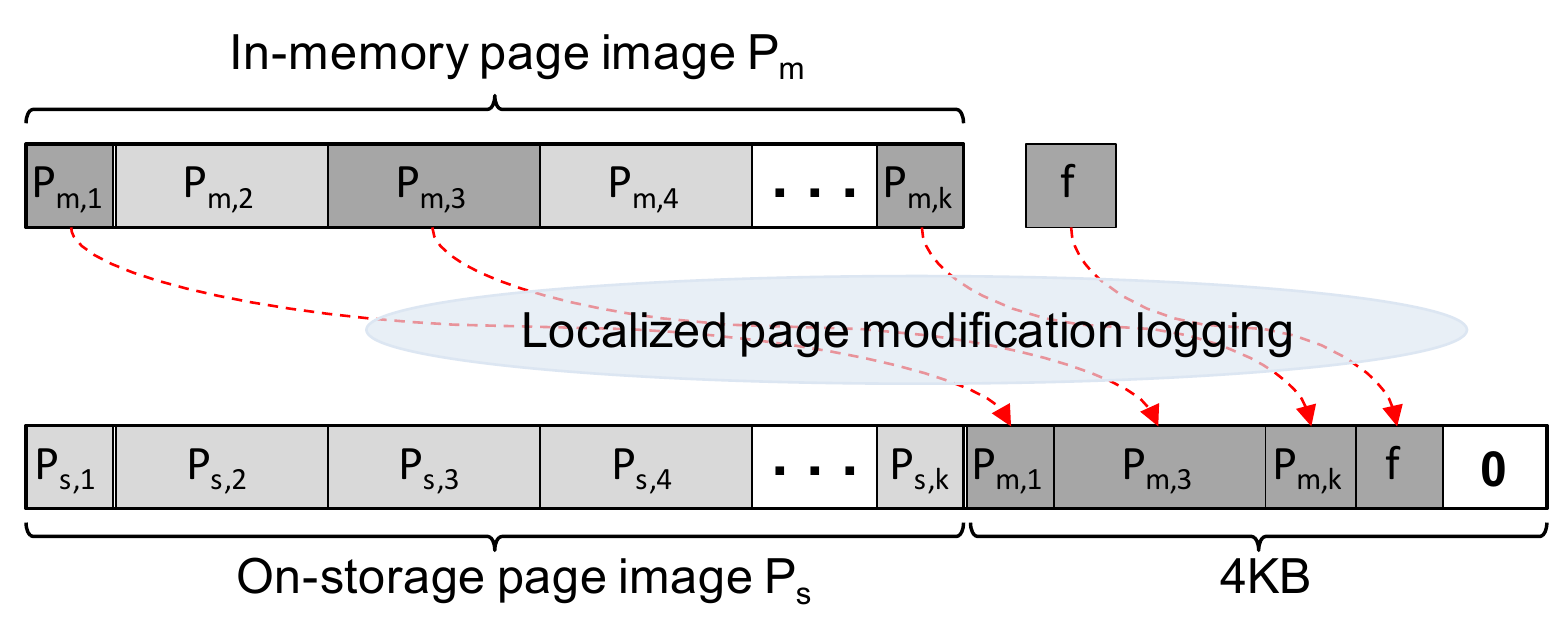}
  \caption{Illustration of the localized page modification logging, where the to-be-flushed in-memory page $P_m$ contains three modified segments $P_{m,1}$, $P_{m,3}$, and $P_{m,k}$.} \label{fig:modificationlogging}
\end{figure}

Fig.~\ref{fig:modificationlogging} further illustrates this implementation strategy.
Among the all the $k$ segments, the first segment $P_{m,1}$ is the page header and the last segment $P_{m,k}$ is the page trailer, both of which can be much smaller than the other segments. Suppose an page update causes modification of the segment $P_{m,3}$ and page header/trailer. When B-tree evicts this page from the memory, it constructs the $\Delta$ as $[P_{m,1}, P_{m,3}, P_{m,k}]$, and writes $\Delta$ and the $k$-bit vector $f$ to the dedicated 4KB block logging block, which is further compressed inside the storage device.

We note that, if B-tree treats in-memory pages as immutable and uses in-memory delta chaining to keep track of the in-memory page modification (which is used in the Bw-tree~\cite{levandoski2013bw, levandoski2013bw-ICDE} to achieve latch-free operations), we can most likely further reduce $|\Delta|$ and hence improve the effectiveness of the localized page modification logging on reducing the write amplification. However, such delta-chaining approach can largely complicate the B-tree implementation~\cite{wang2018building} and incur noticeable memory usage overhead. Hence, this work chooses the above simple intra-page segment-based tracking approach in our implementation and evaluation.  

\subsection{Sparse Redo Logging}
\label{sec:sparselogging}
The third design technique aims at reducing the component $\alpha_{log}$ in Eq.~(\ref{eq:WA})~(i.e., improving the redo log data compressibility). To maximize the reliability, B-tree flushes the redo log with {\it fsync} or {\it fdatasync} at every transaction commit. In order to reduce the log-induced storage overhead, conventional practice always tightly packs log records into the redo log. As a result, multiple consecutive redo log flushes may write to the same LBA block on the storage device, especially when transaction records are significantly smaller than 4KB and/or the workload concurrency is not very high. This can be illustrated in Fig.~\ref{fig:logging}:
\begin{figure}[b]
  \centering
  \includegraphics[width=\linewidth]{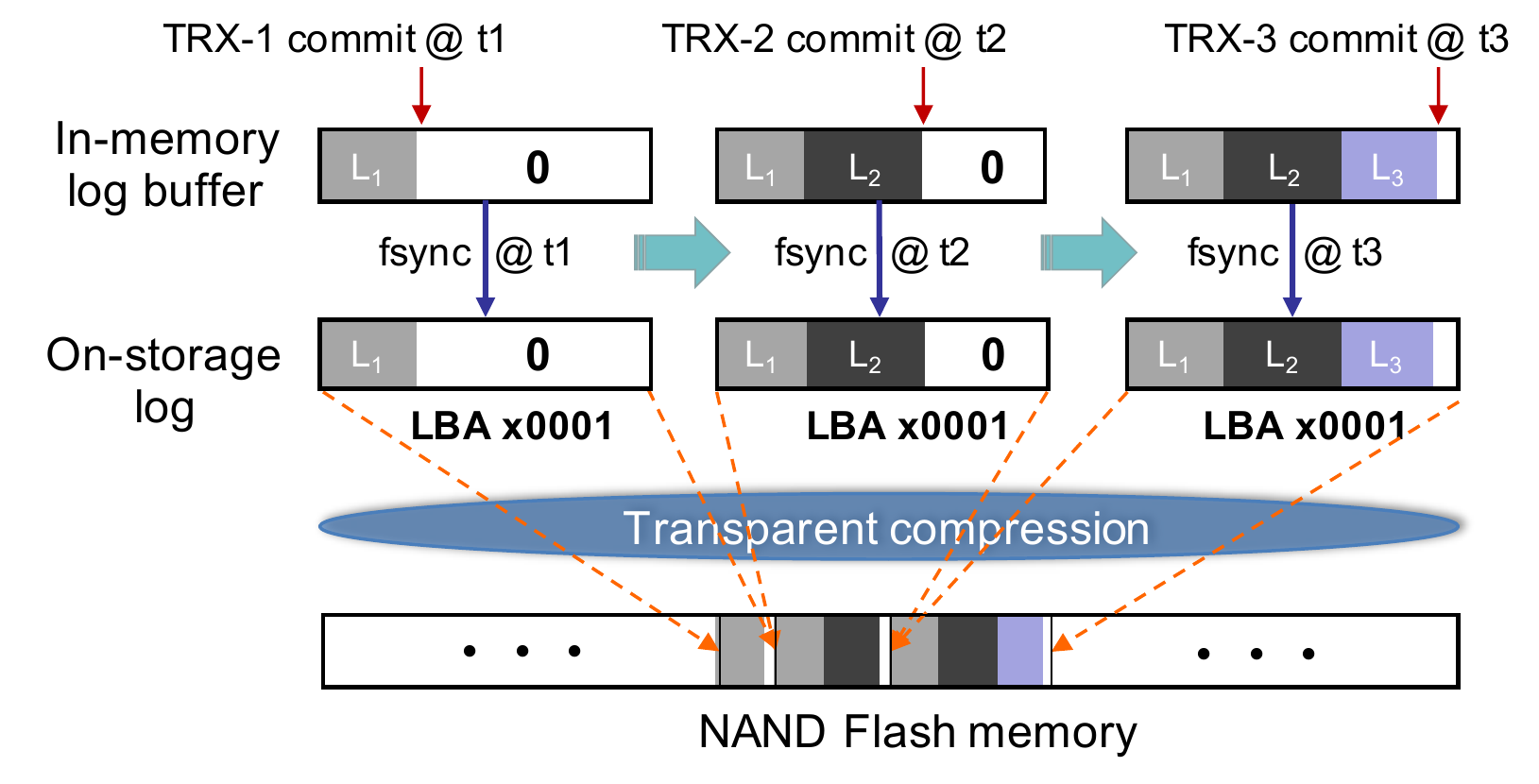}
  \caption{Conventional implementation of redo logging where log records are tightly packed into redo log and consecutive transactions commits could flush redo log to the same LBA (e.g., LBA 0x0001 in this example) multiple times.} \label{fig:logging}
\end{figure}
Suppose three transactions TRX-1, TRX-2, and TRX-3 (with log records $L_1$, $L_2$, and $L_3$) commit at the time $t_1$, $t_2$, and $t_3$, respectively, where $t_1< t_2<t_3$. As illustrated in Fig.~\ref{fig:logging}, at the time $t_1$, 4KB data $[L_1, {\bf O}]$ is flushed from the in-memory redo log buffer to the LBA 0x0001 on the storage device that further internally compresses the data. Later on, the log record $L_2$ is appended into the redo log buffer, and at the time $t_2$, the 4KB data $[L_1, L_2, {\bf O}]$ is flushed to the same LBA 0x0001 on the storage device. Similarly, at the time $t_3$, the 4KB data $[L_1, L_2, L_3, {\bf O}]$ is flushed to the same LBA 0x0001 on the storage device. As illustrated in Fig.~\ref{fig:logging}, the same log record (e.g., $L_1$ and $L_2$) are written to the storage device multiple times, leading to a higher write amplification. Equivalently, as more log records are accumulated inside each 4KB redo log buffer block, the redo log data compression ratio $\alpha_{log}$ will become worse and worse over the multiple consecutive redo log flushes.

\begin{figure}[hbtp]
  \centering
  \includegraphics[width=\linewidth]{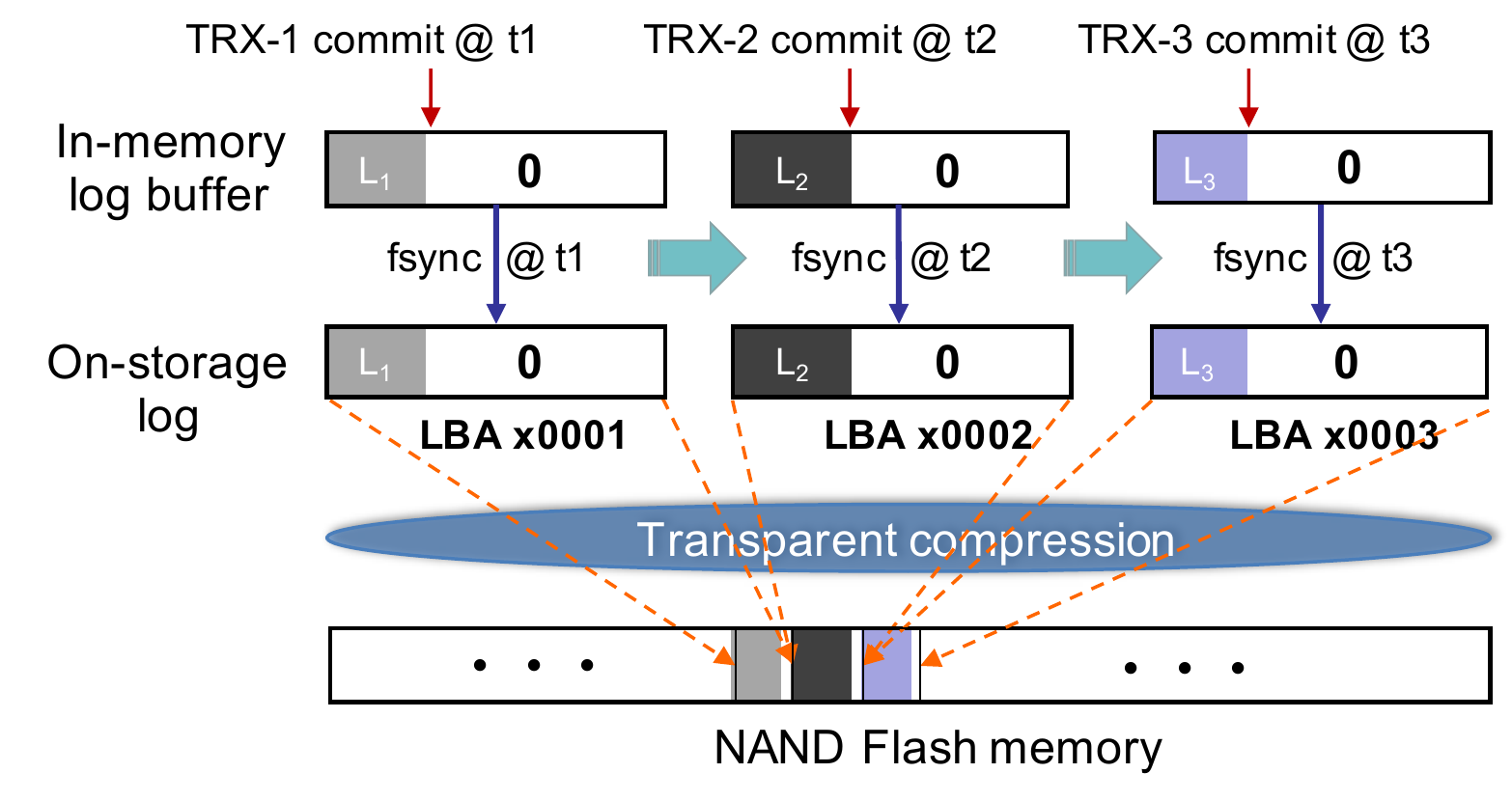}
  \caption{Illustration of the proposed sparse logging where each redo log flush always writes to a new LBA block.} \label{fig:sparselogging}
\end{figure}

By applying sparse data structure enabled by storage hardware with built-in transparent compression, we propose a design technique called sparse redo logging that can enable the storage hardware most effectively compress the redo log  and hence reduce the logging-induced write amplification. Its basic idea is very simple: At each transaction commit and its corresponding redo log memory-to-storage flush, we always pad zeros into the in-memory redo log buffer to make its content 4KB-aligned. As a result, the next log record will be written into a new 4KB space in the redo log buffer. Therefore, each log record will be written to the storage device only once, leading to a lower write amplification compared with the conventional practice. This can be further illustrated in Fig.~\ref{fig:sparselogging}: Assuming the same scenario as shown above in Fig.~\ref{fig:logging}, after the transaction TRX-1 commits at the time $t_1$, we pad zeros into the redo log buffer and flush the 4KB data $[L_1, {\bf O}]$ to the LBA 0x0001 on the storage device. Subsequently, we put the next log record $L_2$ in a new 4KB space in the redo log buffer. At the time $t_2$, the 4KB data $[L_2, {\bf O}]$ is flushed to a new LBA 0x0002 on the storage device. Similarly, at the time $t_3$, the 4KB data $[L_3, {\bf O}]$ is flushed to another new LBA 0x0003 on the storage device. Clearly, each redo log record is written to the storage device only once, and redo log writes can be (much) better compressed by the storage hardware, leading to a (much) smaller $\alpha_{log}$ and hence lower write amplification. Since each transaction commit always invokes one 4KB write to the storage device in both conventional logging and proposed sparse logging, the total redo log write volume $W_{log}$ in Eq.~(\ref{eq:WA}) will remain the same. Therefore, by reducing the log compression ratio $\alpha_{log}$, the proposed sparse logging reduces the component $\alpha_{log}\cdot W_{log}$ in the total B-tree write amplification.

\begin{figure*}[hbtp]
  \centering
  \includegraphics[width=\linewidth]{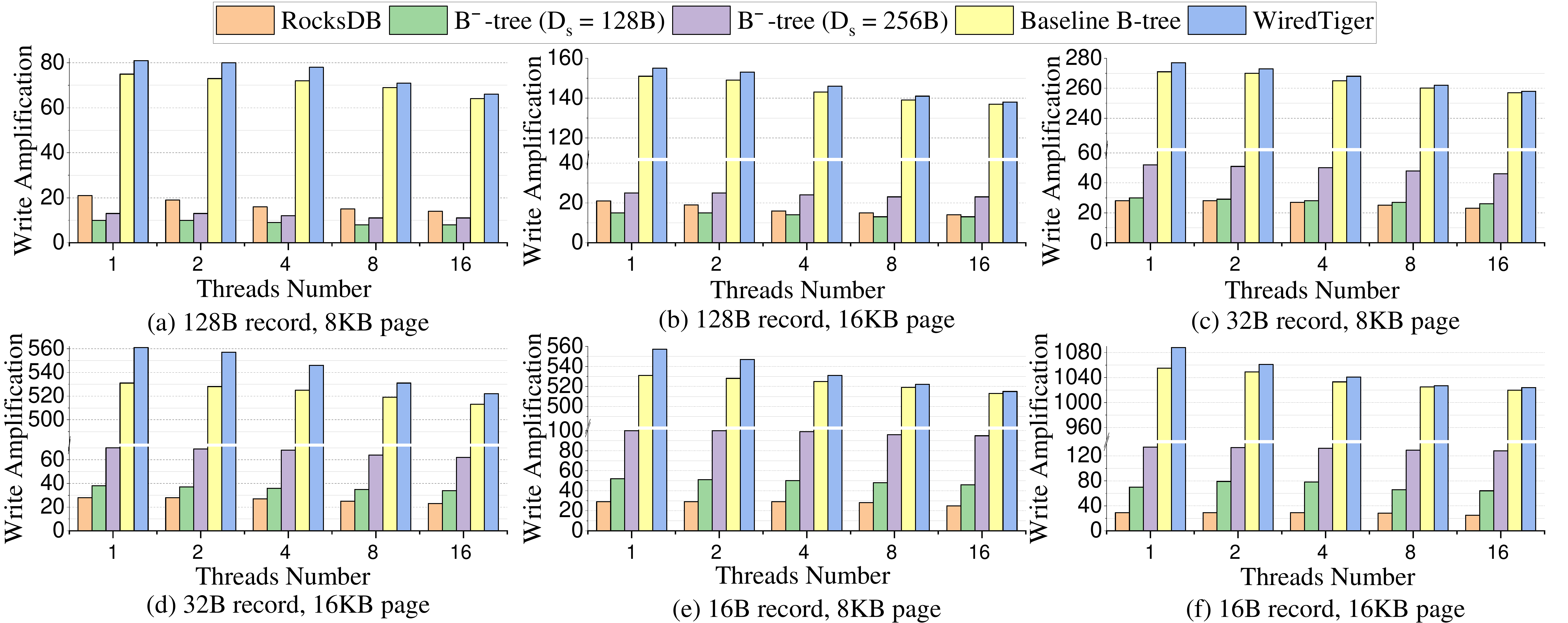}
  \caption{Write amplification under the log-flush-per-minute policy, where the dataset size is 150GB and cache size is 1GB.} \label{fig:WA1}
\end{figure*}

\section{Evaluation}
\label{sec:evaluation}

For the purpose of demonstration and evaluation, we implemented a B-tree (referred to as B$^\shortminus$-tree) that incorporates the above presented three simple design techniques. To facilitate the comparison, we also implemented and evaluated a baseline B-tree that uses the conventional page shadowing, where we persist the page table after each page flush. In fact, since the proposed three design techniques mainly confine within the I/O module and are largely orthogonal to the core B-tree in-memory architecture and operation, we obtained the B$^\shortminus$-tree by simply incorporating the proposed design techniques into the baseline B-tree with 1,200 LoC added/modified. Moreover, we also considered RocksDB and WiredTiger as representatives of LSM-tree and B-tree. For RocksDB, we set its maximum number of compaction and flush threads as 12 and 4, and set the Bloomfilter as 10 bits per record. For WiredTiger and our own baseline B-tree and B$^\shortminus$-tree, we use 4 background write threads that flush dirty in-memory pages to the storage device.

\subsection{Experimental Setup}
\label{sec:setup}
We ran all the experiments on a server with 24-core 2.6GHz Intel CPU, 64GB DDR4 DRAM, and a 3.2TB computational storage drive with built-in transparent compression that was recently launched to the commercial market by ScaleFlux~\cite{ScaleFlux-link}. This 3.2TB drive carries out hardware-based zlib compression on each 4KB block directly along the I/O path. The per-4KB compression/decompression latency of the hardware zlib engine is around 5$\mu$s, which is over 10$\times$ shorter than the TLC/QLC NAND flash memory read latency~($\sim$80$\mu$s and above) and write latency~($\sim$1ms and above). Operating with PCIe Gen3$\times$4 interface, this computational storage drive can achieve up to 3.2GB/s sequential throughput and 650K (520K) random 4KB read~(write) IOPS~(I/O per second) over 100\% LBA span. In comparison, leading-edge commodity NVMe SSDs (e.g., Intel P4610) achieve similar sequential throughput and random 4KB read IOPS, but have much worse random 4KB write IOPS (e.g., below 300K). This is because built-in transparent compression can significantly reduce the garbage collection overhead inside the storage drive.  

This 3.2TB computational storage drive can report the amount of post-compression data being physically written to the NAND flash memory inside the drive, which are used in the calculation of write amplification. Before measuring the write amplification for each case, we populate the B-tree/LSM-tree data store by inserting all the data records in a fully random order. Once after the data store has been fully populated, we subsequently run random write-only workloads over one hour in order to measure the write amplification. In all our experiments, we generate the content of each record by filling its half content as all-zero and the other half content as random bytes in order to mimic the runtime data content compressibility. 

We note that the effectiveness of the proposed sparse redo logging strongly depends on the redo log flush policy. As discussed above Section~\ref{sec:sparselogging}, when redo log flushes at every transaction commit to maximize the system reliability, sparse redo logging is very effective. However, for applications that can tolerate the loss of certain amount of most recent data, one could relax the redo log flush policy (e.g., flush every one minute) under which the proposed sparse redo logging will be much less useful. Therefore, we considered two scenarios in our evaluation: (1) redo log flush per transaction commit (denoted as {\it log-flush-per-commit}), and (2) redo log flush per minute (denoted as {\it log-flush-per-minute}). 

\begin{figure*}[hbtp]
  \centering
  \includegraphics[width=\linewidth]{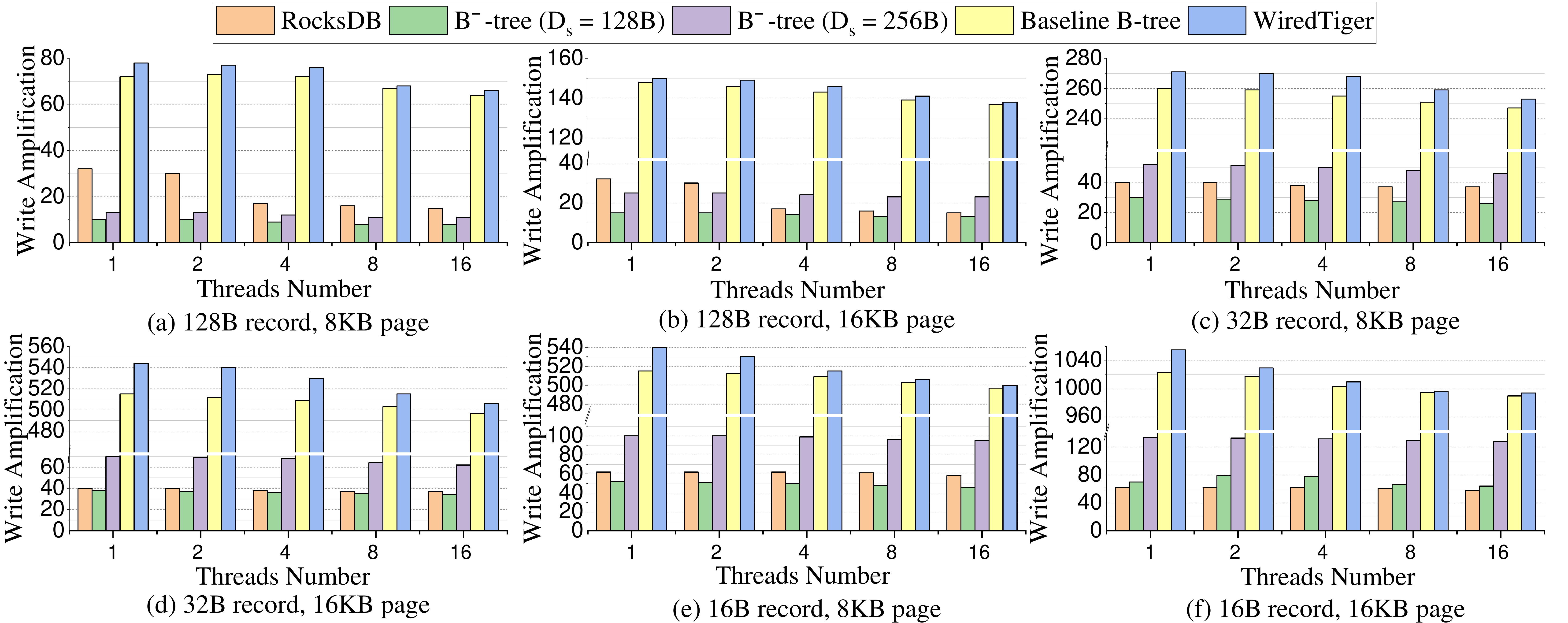}
  \caption{Write amplification under the log-flush-per-minute policy, where the dataset size is 500GB and cache size is 15GB.} \label{fig:WA2}
\end{figure*}

\subsection{Experiments with Log-Flush-Per-Minute}
\label{sec:wawithoutlog}
We first carried out experiments without taking into account of the benefit of sparse redo logging (i.e., setting the redo log flush policy as per-minute). We considered two different dataset size: (1) 150GB dataset with 1GB cache memory, and (2) 500GB dataset with 15GB cache memory. We also considered three different record size~(including 8B key): 128B, 32B, and 16B. For B-tree implementations, following the popular RDBMs such as Oracle and MySQL, we considered two different page size, including 8KB and 16KB. For our B$^\shortminus$-tree, the implementation of the page modification logging involves the following two parameters: (1) the threshold $T$ that determines the maximum $|\Delta|$ per page, and (2) the segment size~(denoted as $D_s$) when partitioning each page into multiple segments for tracking page modification, as discussed in Section~\ref{sec:pagelogging}. 

Fig.~\ref{fig:WA1} and Fig.~\ref{fig:WA2} show the measured write amplification for 150GB and 500GB datasets, respectively. In each experiment, we use either 1, 2, 4, 8, or 16 client threads to cover a wide range of workload concurrency. For B$^\shortminus$-tree, we set the threshold $T$ as 2KB, and set the segment size $D_s$ as either 128B or 256B. Since both WiredTiger and our own baseline B-tree uses page shadowing, they have very similar write amplification as shown in Fig.~\ref{fig:WA1} and Fig.~\ref{fig:WA2}. Compared with RocksDB, normal B-tree (i.e., WiredTiger and our own baseline B-tree) has a much larger write amplification, while our B$^\shortminus$-tree can essentially close the B-tree vs.~LSM-tree write amplification gap. For example, in the case of 500GB dataset and 32B record size and 4 client threads, the write amplification of RocksDB is 38, while the write amplification of WiredTiger is 268 under 8KB page size and 530 under 16KB page size, respectively, which are 7.1$\times$ and 13.9$\times$ larger than that of RocksDB. In comparison, the write amplification of B$^\shortminus$-tree with $D_s$=128B is 28 under 8KB page size (which is only 73.7\% of RocksDB's write amplification) and 36 under 16KB page size (which is almost the same as RocksDB). 

As shown in both Fig.~\ref{fig:WA1} and Fig.~\ref{fig:WA2}, the write amplification of both normal B-tree and B$^\shortminus$-tree will increase as we reduce the record size (e.g., from 128B per record to 16B per record) and/or increase the B-tree page size (i.e., from 8KB to 16KB). Since we use the log-flush-per-minute policy, the overall write amplification of both normal B-tree and B$^\shortminus$-tree tends to be dominated by the $\alpha_{pg}\cdot WA_{pg}$, as shown in Eq.~(\ref{eq:WA}). 
In the case of normal B-tree, $WA_{pg}$ proportionally increases as we reduce the record size and/or increase the page size. Therefore, the write amplification of normal B-tree almost linearly scale with the page size and the inverse of the record size. In the case of B$^\shortminus$-tree, its $\alpha_{pg}\cdot WA_{pg}$ not only depends on the record size and page size, but also depends on the threshold $T$ and segment size $D_s$. Hence, the write amplification of B$^\shortminus$-tree tends to sub-linearly scale with the page size and the inverse of the record size, as shown in both Fig.~\ref{fig:WA1} and Fig.~\ref{fig:WA2}. In contrast, the write amplification of RocksDB is weakly dependent on the record size. 

As the number of client threads increases, the write amplification of normal B-tree noticeably reduces, because of the larger probability of page flush coalescing under higher workload concurrency. In comparison, the write amplification of B$^\shortminus$-tree is much more weakly dependent on the number of client threads, because the probability that different client threads modify the same segment inside a page is much smaller than the probability that different client threads modify the same page. Moreover, the write amplification of B$^\shortminus$-tree increases as we increase the segment size $D_s$, simply because the page modification logging is done in the unit of segments. The impact of segment size $D_s$ is more significant under smaller record size, as shown in both Fig.~\ref{fig:WA1} and Fig.~\ref{fig:WA2}.
\begin{figure*}[hbtp]
  \centering
  \includegraphics[width=\linewidth]{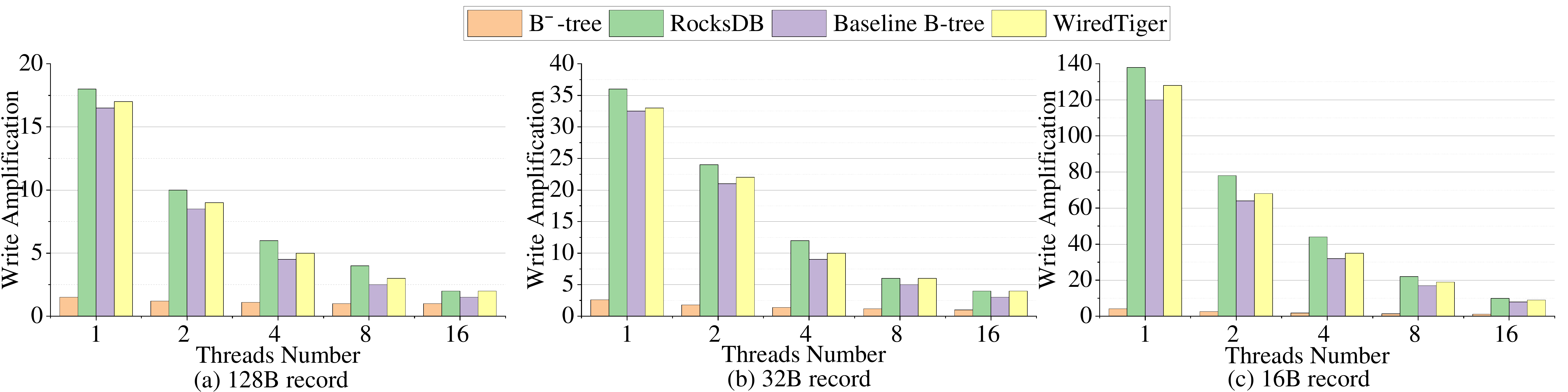}
  \caption{Log-induced write amplification when using the log-flush-per-commit policy.} \label{fig:WA3}
\end{figure*}

The write amplification of LSM-tree may noticeably increase as the dataset size increases, which can be observed by comparing the results in Fig.~\ref{fig:WA1} and Fig.~\ref{fig:WA2}. This is because a larger dataset size results in more levels in LSM-tree, while the write amplification of LSM-tree tends to increase with the number of levels. In contrast, the write amplification of B-tree is very weakly dependent on the dataset size. As a result, the write amplification comparison of RocksDB vs.~B$^\shortminus$-tree is noticeably different between the 150GB dataset and 500GB dataset. In the case of 150GB dataset as shown in Fig.~\ref{fig:WA1}, the write amplification of RocksDB can be up to $2\times$ larger than that of B$^\shortminus$-tree (under 128B per record and 8KB page size), and can be up to $4\times$ smaller than that of B$^\shortminus$-tree (under 16B per record and 16KB page size). In comparison, in the case of 500GB dataset as shown in Fig.~\ref{fig:WA2}, the write amplification of RocksDB can be up to $3\times$ larger than that of B$^\shortminus$-tree (under 128B per record and 8KB page size), and can be up to $2\times$ smaller than that of B$^\shortminus$-tree (under 16B per record and 16KB page size). The results clearly show that, even without taking into account of the effectiveness of sparse redo logging, the proposed B$^\shortminus$-tree can already close the write amplification gap between B-tree and LSM-tree.

\subsection{Experiments with Log-Flush-Per-Commit}
\label{sec:wawithlog}
We carried out further experiments by switching to the log-flush-per-commit policy, under which the proposed sparse redo logging can noticeably contribute to reducing the write amplification. First, Fig.~\ref{fig:WA3} shows the measured write amplification caused by the log flush, i.e., the $\alpha_{log}\cdot WA_{log}$ component in Eq.~\ref{eq:WA}. Given the record size, except the case of B$^\shortminus$-tree, the log-induced write amplification significantly reduces as we increase the number of client threads. This is because, under higher workload concurrency, more transaction commits can be coalesced in each log flush. In contrast, the log-induced write amplification of B$^\shortminus$-tree is much weakly dependent on the number of client threads, because of its use of sparse redo logging. As the record size reduces, the log-induced write amplification almost proportionally increases when not using the sparse redo logging. The results in Fig.~\ref{fig:WA3} clearly demonstrate the effectiveness of the proposed sparse redo logging design technique when data management systems use the log-flush-per-commit policy to improve the data reliability. 
\begin{figure*}[hbtp]
  \centering
  \includegraphics[width=\linewidth]{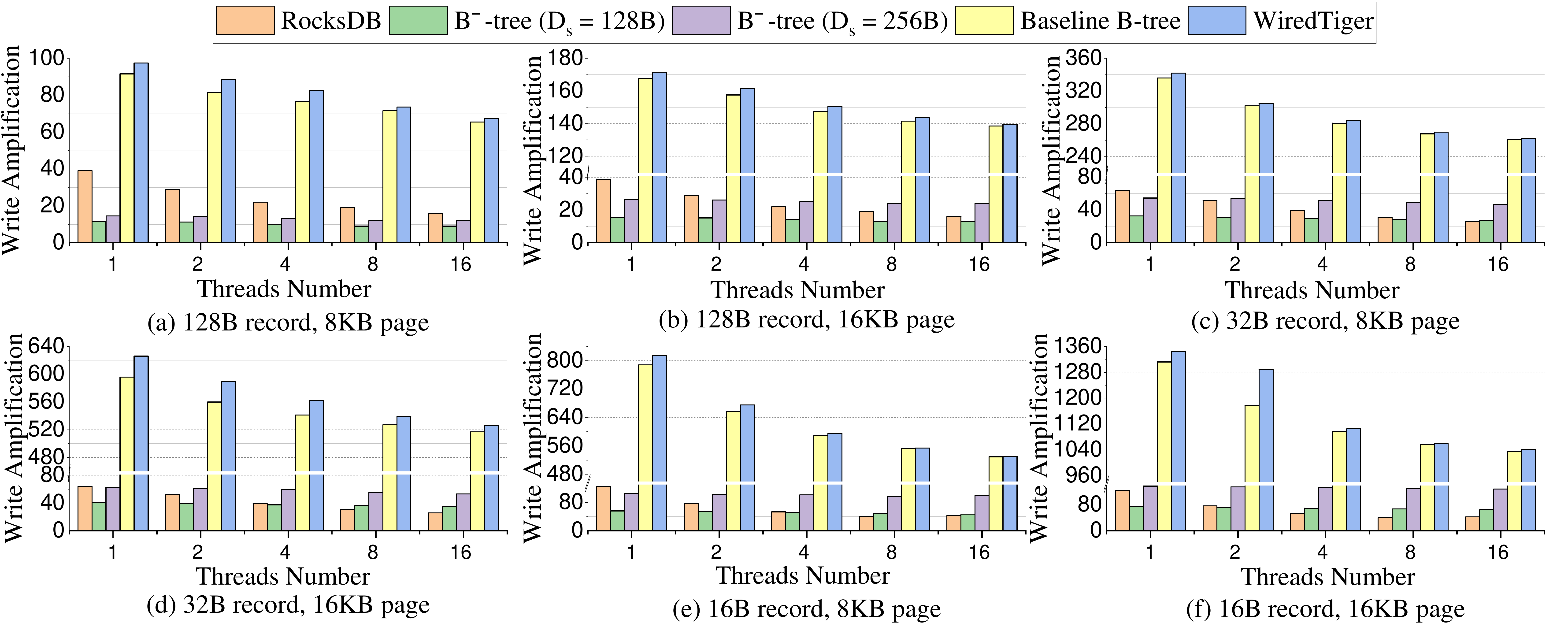}
  \caption{Write amplification under the log-flush-per-commit policy, where the dataset size is 150GB and cache size is 1GB.} \label{fig:WA4}
\end{figure*}


Fig.~\ref{fig:WA4} further shows the total write amplification under the log-flush-per-commit policy, where the dataset size is 150GB and cache size is 1GB. Compared with the experiments under the log-flush-per-minute policy (as shown in Fig.~\ref{fig:WA1}), the write amplification of B$^\shortminus$-tree remains almost the same, while the write amplification of the other three cases (i.e., RocksDB, our own baseline B-tree, and WiredTiger) noticeably increases, especially when the number of client threads is small. As a result, B$^\shortminus$-tree can more effectively close the B-tree vs.~LSM-tree write amplification gap and achieve better-than-RocksDB write amplification under more scenarios.

\subsection{Impact of Threshold $T$}
As discussed earlier in Section~\ref{sec:pagelogging}, the proposed page modification logging design approach is subject to a write amplification vs.~storage usage trade-off that is configured by the threshold $T\in(0, 4\textrm{KB}]$.  As we increase the value of $T$, we can pack more modification logs into each dedicated 4KB log space in order to further reduce the total write amplification, which nevertheless meanwhile induces higher storage usage overhead. All the experiments above were carried out with $T$ as 2KB. We carried out experiments under different values of threshold $T$ to study its impact on the write amplification vs.~storage usage trade-off. For each B-tree page $P_i$, let $|\Delta_i|$ denote the size of its associated modification log. Let $N$ denote the total number of B-tree pages and recall that $l_{pg}$ denotes the page size, we can express the average storage usage overhead factor as
\begin{equation}
    \beta = \frac{\sum_{i=1}^N |\Delta_i|}{N\cdot l_{pg}}.
\end{equation}
Under sufficiently large $N$, the value of $\beta$ mainly depends on the page size $l_{pg}$, the threshold $T$, and the workload characteristics~(in particular the write request distribution over all the pages). It also weakly depends on the segment size $D_s$. Assuming the fully random write request distribution, we carried out experiments to measure the average value of $\beta$, and the results are summarized in Table~\ref{tab:sc}. The results clearly show that the storage usage overhead will reduce as we reduce the threshold $T$ and/or increase the page size. The impact of the segment size $D_s$ is very marginal.

\begin{table}[htbp]
\centering
\caption{Storage usage overhead factor $\beta$ of B$^\shortminus$-tree.}
\label{tab:sc} 
\begin{tabular}{|c|c|c|c|c|}
\hline
\multirow{ 2}{*}{Page size} & \multirow{ 2}{*}{$D_s$} & \multicolumn{3}{c|}{Threshold $T$}\\
\cline{3-5}
& & 4KB & 2KB & 1KB \\
\hline
\multirow{ 2}{*}{8KB} & \hspace*{3pt} 128B \hspace*{3pt} & \hspace*{3pt}27.0\% \hspace*{3pt}& \hspace*{3pt}12.4\% \hspace*{3pt}& \hspace*{3pt} 5.6\% \hspace*{3pt}\\ \cline{2-5}
 & 256B & 26.3\% & 11.5\% & 4.8\% \\
\hline
\multirow{ 2}{*}{16KB} & 128B & 12.7\% & 6.0\% & 2.8\%
\\\cline{2-5}
& 256B & 12.3\% & 5.6\% & 2.3\% \\
\hline
\end{tabular}
\end{table}

\begin{figure}[htbp]
  \centering
  \includegraphics[width=\linewidth]{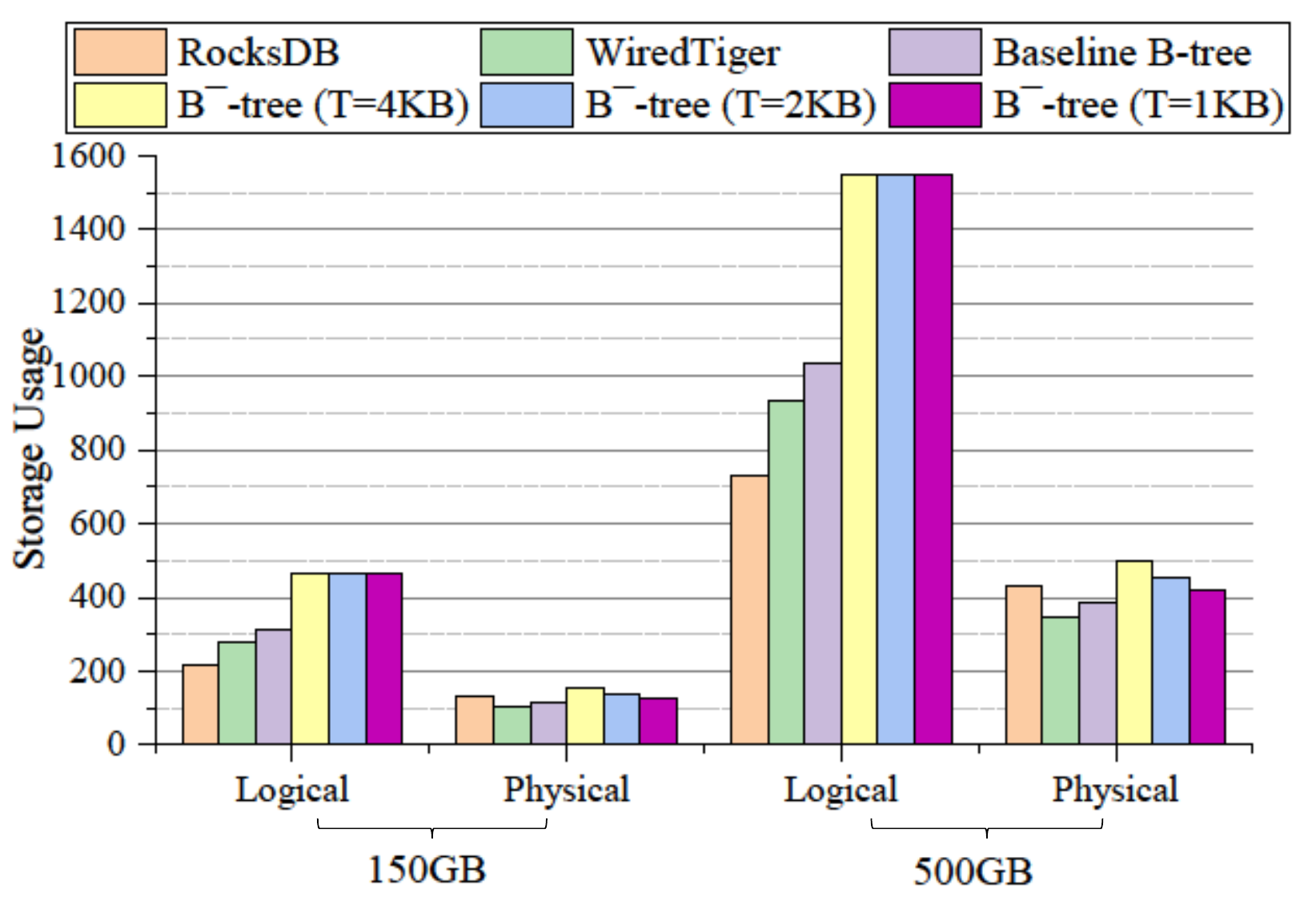}
  \caption{Comparison of logical and physical storage space usage where B-tree page size is 8KB.} \label{fig:datasizecompare}
\end{figure}

Fig.~\ref{fig:datasizecompare} further compares the total storage usage in terms of both logical storage usage on the LBA space (i.e., before in-storage compression) and physical usage of flash memory (i.e., after in-storage compression). Since LSM-tree has a more compact data structure than B-tree, RocksDB has a (much) smaller logical storage usage than the others as shown in Fig.~\ref{fig:datasizecompare}. Since B$^\shortminus$-tree allocates one 4KB block for each page in order to implement the localized modification logging, its logical storage usage is much larger than the normal B-tree. Nevertheless, after the in-storage compression, WiredTiger and our baseline B-tree consume less physical flash memory capacity than RocksDB (because of the space amplification of LSM-tree) and B$^\shortminus$-tree (because of the storage overhead caused by page modification logging). Due to the storage space overhead, B$^\shortminus$-tree has slightly larger physical storage usage than RocksDB. For example, in the case of 500GB dataset size, the physical storage usage of RocksDB is 431GB, while the physical storage usage of B$^\shortminus$-tree with $T$=2KB is 452GB, only about 5\% larger than RocksDB.

\begin{figure}[htbp]
  \centering
  \includegraphics[width=\linewidth]{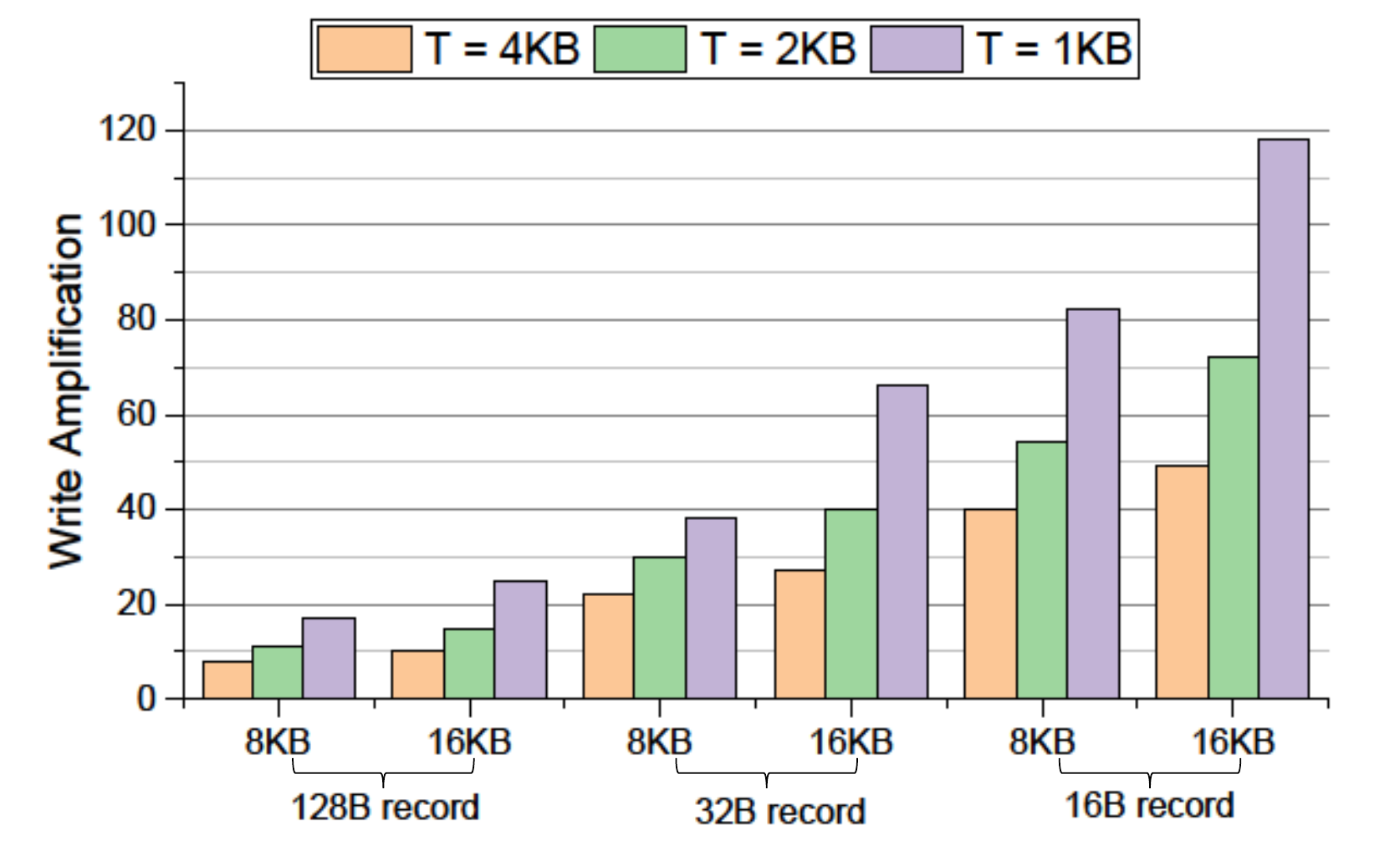}
  \caption{B$^\shortminus$-tree write amplification under different $T$.} \label{fig:delta compare}
\end{figure}

Fig.~\ref{fig:delta compare} compares the write amplification of B$^\shortminus$-tree under different value of the threshold $T$, where we use the log-flush-per-minute policy in order to better show the impact of $T$. The segment size $D_s$ is 128B. The results clearly show that we can reduce the write amplification by increasing the threshold $T$. Moreover, the reduction on the write amplification tends to become less as we continue to increase the threshold $T$. This is because, as the page modification log size $|\Delta|$ becomes larger, the write amplification caused by flushing the modification log becomes larger. Combining the results shown in  Fig.~\ref{fig:datasizecompare} and Fig.~\ref{fig:delta compare}, we can clearly observe the impact of the threshold $T$ on the trade-off between the write amplification and storage usage overhead. The setting of $T$=2KB appears to achieve a reasonable balance and hence has been used in all the experiments presented above in Sections~\ref{sec:wawithoutlog} and \ref{sec:wawithlog}.

\subsection{Speed Performance Evaluation}
\begin{figure}[htbp]
  \centering
  \includegraphics[width=\linewidth]{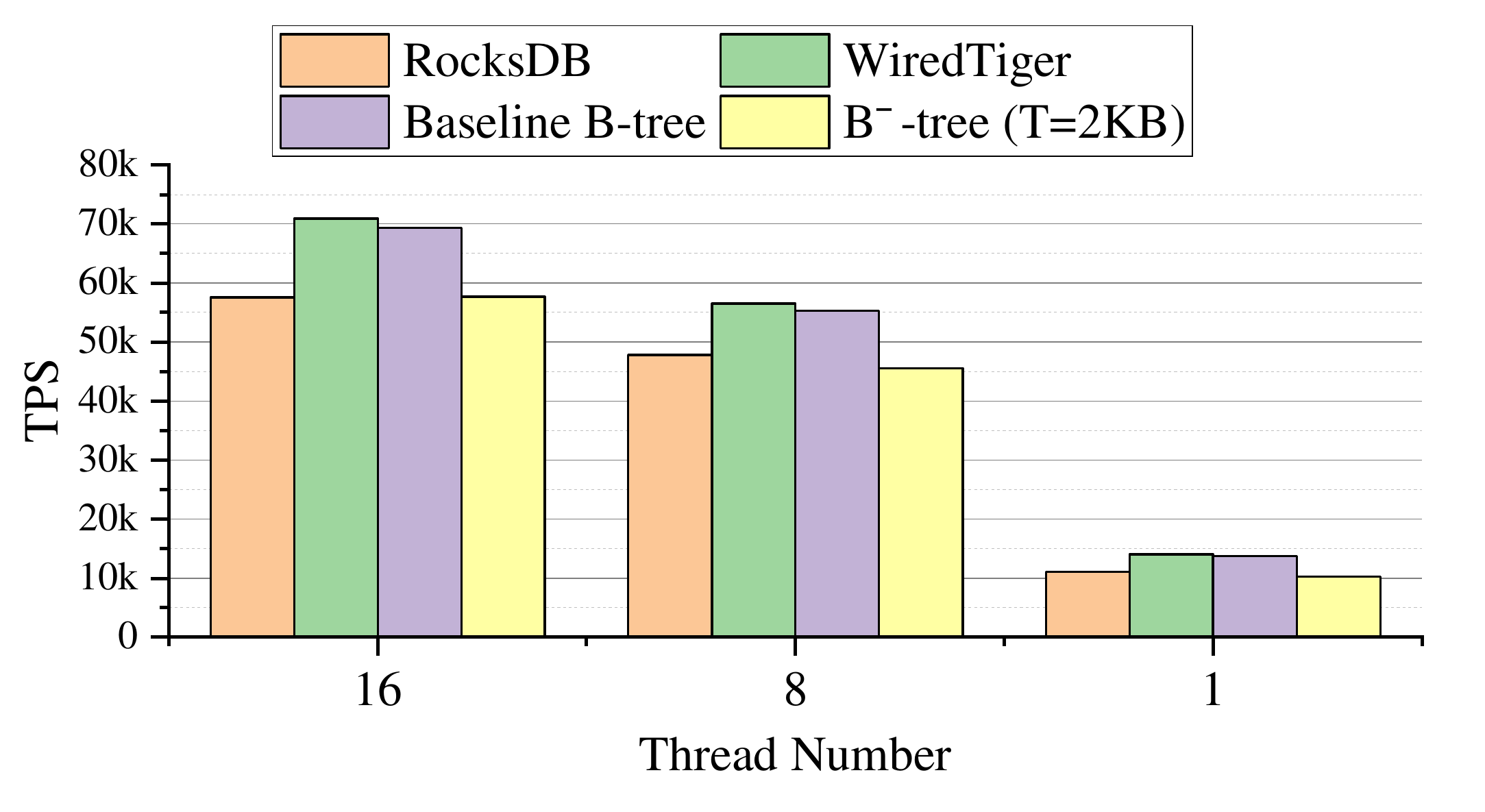}
  \caption{Random point read speed performance measured on 150GB dataset with 1GB cache and 128B per record.} \label{fig:pointreadtps}
\end{figure}

Finally, we studied the speed performance of B$^\shortminus$-tree. Compared with normal B-tree, B$^\shortminus$-tree tends to have lower read speed performance because of the following two overhead when fetching each page from the storage: (1) B$^\shortminus$-tree has to fetch an extra 4KB block from the storage, and (2) B$^\shortminus$-tree has to consolidate the modification log with the current on-storage page image in order to construct the up-to-date in-memory page image. Using the 150GB dataset with 128B per record as the test vehicle, we run random read-only workloads with either point read or range scan queries. The B-tree page size is 8KB in all the experiments. Fig.~\ref{fig:pointreadtps} shows the measured TPS performance under random point read queries. The results show that normal B-tree (WiredTiger and our own baseline B-tree) have the best point read throughput performance. RocksDB and B$^\shortminus$-tree achieve almost the same random point read throughput performance. By using the Bloomfilter, RocksDB almost completely obviates the read amplification problem of classical LSM-tree. Nevertheless, when serving read requests, RocksDB still has to search the memtable and check the Bloomfilter. As shown in  Fig.~\ref{fig:pointreadtps}, the point read throughput gap between normal B-tree and RocksDB/B$^\shortminus$-tree is not significant. For example, under 16 client threads, WiredTiger can achieve 71K TPS, while RocksDB/B$^\shortminus$-tree can achieve 57K TPS, about 19.7\% less than that of WiredTiger.

\begin{figure}[htbp]
  \centering
  \includegraphics[width=\linewidth]{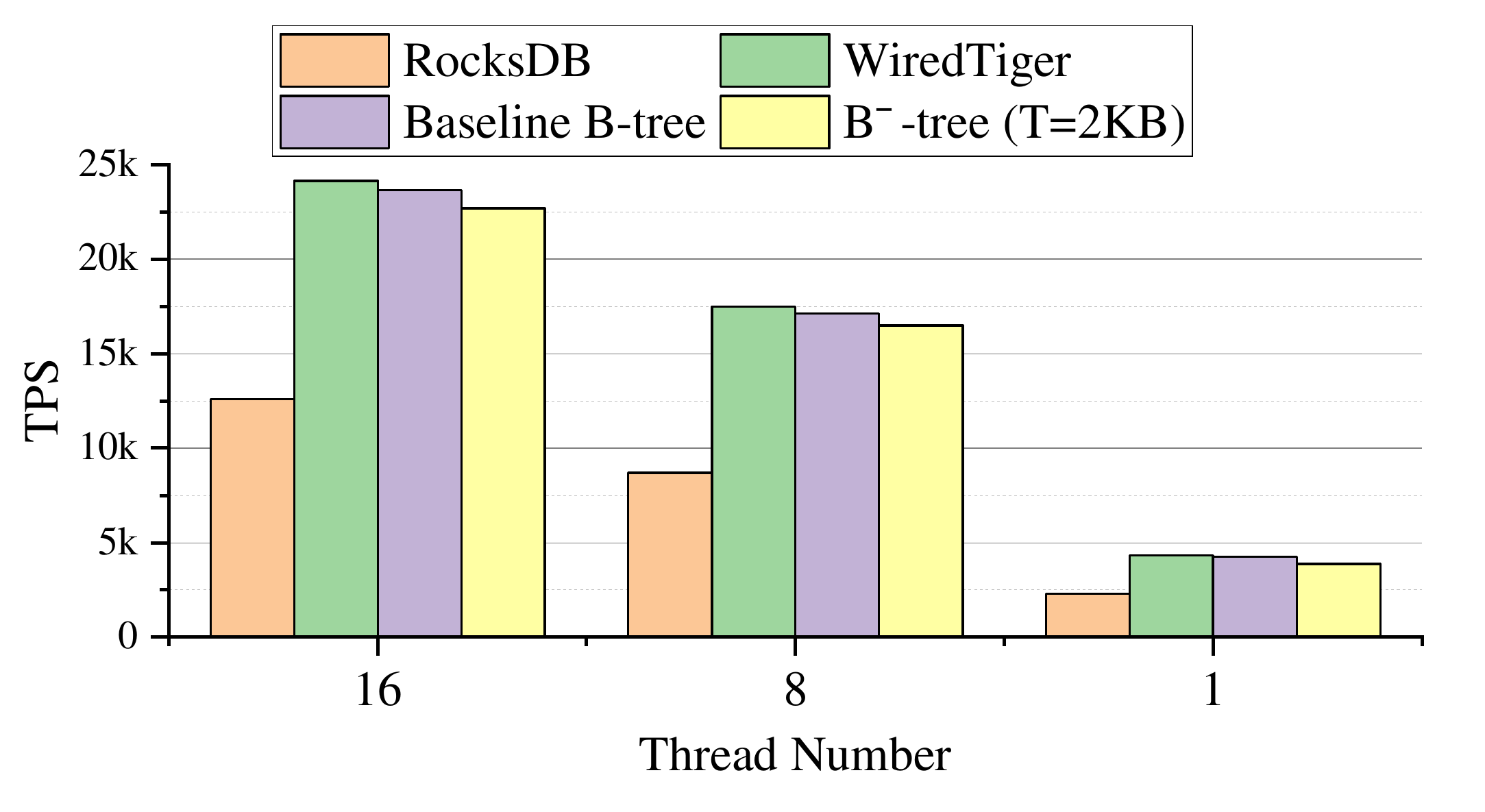}
  \caption{Random range scan speed performance measured on 150GB dataset with 1GB cache and 128B per record, where each range scan covers 100 consecutive records.} \label{fig:readtps}
\end{figure}

Fig.~\ref{fig:readtps} shows the measured TPS when running random range scan queries, where each range scan covers 100 consecutive records. Compared with the case of random point reads, the normal B-tree and B$^\shortminus$-tree have noticeably smaller difference in terms of range scan throughput performance. This is because the two overheads of B$^\shortminus$-tree (i.e., fetching an extra 4KB, and in-memory page reconstruction) can be amortized among the records covered by each range scan. In comparison, RocksDB has noticeably worse range scan throughput performance than the others. This is because range scan invokes (much) larger read amplification on LSM-tree compared with B-tree. 

\begin{figure}[htbp]
  \centering
  \includegraphics[width=\linewidth]{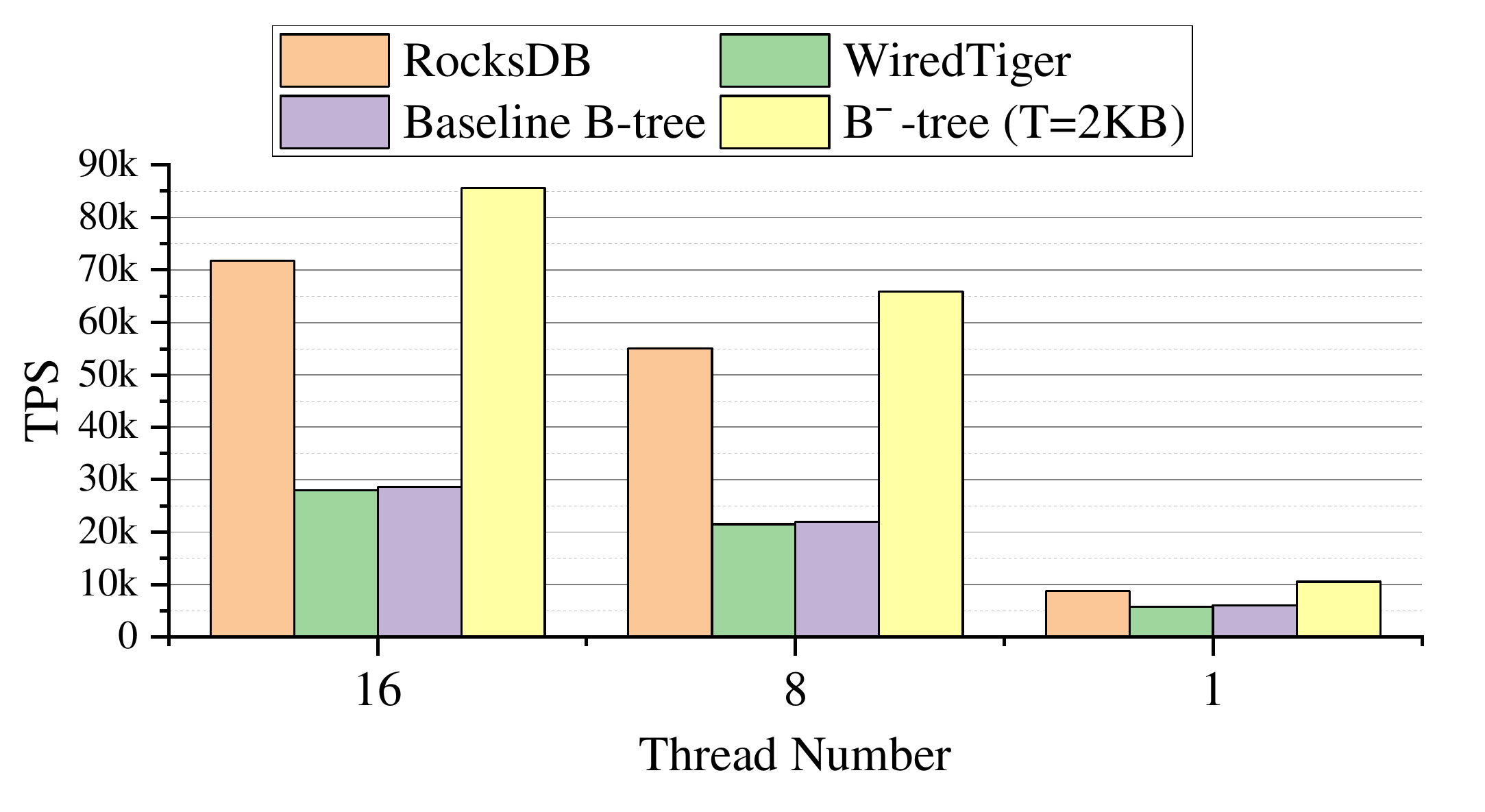}
  \caption{Random write speed performance measured on 150GB dataset with 1GB cache and 128B per record.} \label{fig:tps}
\end{figure}

We also studied the speed performance under random write-only workloads. The random write speed performance of B-tree and LSM-tree is fundamentally limited by the write amplification. Therefore, by significantly reducing the write amplification, B$^\shortminus$-tree should be able to achieve much higher write speed performance. Fig.~\ref{fig:tps} shows the measured random write TPS on 150GB dataset with 128B per record, where the B-tree page size is 8KB. We set the log-flush-per-minute policy in the experiments. Even without the help of the sparse redo logging, B$^\shortminus$-tree achieves achieve 19\% higher write throughput than RocksDB, and about 2.1$\times$ higher write throughput than WiredTiger and our baseline B-tree. The random write speed results well correlate with the write amplification results shown above in Fig.~\ref{fig:WA1}.

\section{Related Work}
\label{sec:related}
Graefe~\cite{graefe2006b} surveyed a variety of design techniques (e.g., I/O optimization, buffering, and relaxing transaction guarantee) that can improve the B-tree write throughput, some of which can reduce the B-tree write amplification. Nevertheless, I/O optimization techniques that mainly aim at converting random page writes to sequential page writes are only useful to HDDs, since modern SSDs achieve almost the same random vs.~sequential write speed performance. Many techniques surveyed in~\cite{graefe2006b} (e.g.,  buffering, relaxing transaction guarantee) are orthogonal to the solutions presented in this paper, and hence can be applied altogether to further reduce the B-tree write amplification. Moreover, copy-on-write or page shadowing~\cite{agrawal1985integrated, kent1985experimental} is a well-known technique to achieve B-tree data atomicity and durability. Compared with B-tree using in-place update, it can reduce the write amplification by about $2\times$. 

Levandoski {\it et al.}~\cite{levandoski2013bw, levandoski2013bw-ICDE} proposed the Bw-tree that can better adapt to modern multi-core CPU architecture and meanwhile reduce the write amplification. Bw-tree treats each in-memory page as immutable and uses delta chaining to keep track of the changes made to each page. This can enable latch-free operations and hence better utilize multi-core CPUs. Meanwhile, by only flushing the delta records, Bw-tree can reduce the write amplification. Bw-tree uses a log-structured store to persist all the pages and deltas, which however suffers from read amplification and background garbage collection overheads. When running Bw-tree on storage hardware with build-in transparent compression, one could enhance Bw-tree by replacing the log-structured store with the localized page modification logging presented in this work. 

B$^\varepsilon$-tree~\cite{brodal2003lower} is another variant of B-tree that can significantly reduce the write amplification through data buffering at non-leaf nodes. It has been used in the design of filesystem~\cite{esmet2012tokufs, jannen2015betrfs, jannen2015betrfs-TOS, yuan2016optimizing} and key-value store~\cite{papagiannis2016tucana, conway2020splinterdb}. In essence, B$^\varepsilon$-tree cleverly mixes the key design principles of B-tree and LSM-tree. Similar to LSM-tree, B$^\varepsilon$-tree has worse range scan speed performance than B-tree. Percona TokuDB~\cite{TokuDB-link} is one publicly known database product that is built upon B$^\varepsilon$-tree.

Little prior research has been done on studying how data management systems could take advantage of storage hardware with built-in transparent compression. Recently, Zheng {\it et al.}~\cite{Zheng-Hotstorage-20} discussed some possible options on leveraging such modern storage hardware to improve data management software design. Chen {\it et al.}~\cite{chen2021kallaxdb} presented a hash-based key-value store that can leverage such modern storage hardware to obviate the use of costly in-memory hash table.

\section{Conclusions}
\label{sec:conclusions}
This paper presents three simple yet effective design techniques that enable B-tree take significant advantage of modern storage hardware with built-in transparent compression. By decoupling logical vs.~physical storage space utilization efficiency, such new storage hardware allows data management systems employ sparse data structure without sacrificing the true physical data storage cost. This opens a new but largely unexplored spectrum of opportunities to innovate data management system design. As one small step towards exploring this design spectrum, this paper presents three design techniques that can appropriately embed sparsity into B-tree  data structure to largely reduce the B-tree write amplification. Experimental results show that the proposed design techniques can reduce the B-tree write amplification by over 10$\times$, which essentially closes the B-tree vs.~LSM-tree gap in terms of write amplification. This work suggests that the arrival of such new storage hardware warrants a revisit on the role and comparison of B-tree and LSM-tree in future data management systems.

\bibliographystyle{ACM-Reference-Format}
\bibliography{reference}
\end{document}